\documentstyle[psfig]{l-aa}

\begin{document}

   \thesaurus{03
              (11.01.2;
               11.09.1;
               11.11.1)}

   \title{A radio galaxy at z=3.6 in a giant rotating
Lyman $\alpha$ halo\thanks{Based on observations collected at the European 
Southern Observatory,
La Silla, Chile, and at the Very Large Array. The VLA is a facility of the
U.S. National Radio Astronomy Observatory, which  is operated by Associated
Universities, Inc., under  cooperative agreement with the U.S. National Science
Foundation}}

   \author{R. van Ojik \inst{1}, H.J.A. R\"ottgering \inst{1,2,3}
           C.L. Carilli \inst{1,4} G.K. Miley \inst{1} M.N. Bremer \inst{1}
           F. Macchetto \inst{5,6}
          }
   \offprints{H.J.A. R\"ottgering}

\institute{Leiden Observatory, P.O. Box 9513, 2300 RA, Leiden,
The Netherlands \and
Mullard Radio Astronomy Observatory, Cavendish Laboratory,
Cambridge CB3 0HE, England \and 
Institute of Astronomy, Madingley Road, Cambridge CB3 0HA, England \and
Harvard-Smithsonian Center for Astrophysics, Cambridge, MA 02138, USA \and
Space Telescope Science Institute, 3700 San Martin Drive, Baltimore MD21218, USA \and
Affiliated with the Space Science Department of ESA
             }

\date{}

\maketitle
\markboth{van Ojik et al.; A radio galaxy at z=3.6 in a giant rotating Lyman
$\alpha$ halo}{}

\begin{abstract}
We present the discovery and detailed observations of the radio galaxy 1243+036
(=4C\,03.24) at a redshift of $z=3.57$.
The radio source was selected on the basis of its extremely steep radio 
spectrum, suggesting that it might be very distant.
The radio source was identified with a galaxy of $R$ magnitude 22.5.
Subsequent spectroscopy showed strong Lyman $\alpha$ and [O\,III] emission,
indicating that the object is a radio galaxy at $z=3.57$.
High resolution ($0.2''$) radio maps show an FRII type radio source with
a sharply bent radio structure. Strong depolarization of the radio emission
indicates that the source is embedded in a magneto-ionic medium. 

The most spectacular feature of 1243+036 is the presence of a Ly$\alpha$ halo
of luminosity $\sim10^{44.5}$ ergs s$^{-1}$ which extends over$\sim20''$ 
(135 kpc).
A $0.6''$ resolution Ly$\alpha$ image shows that the emission line gas is
aligned with main axis of the radio source and has structure down to the 
scale of the resolution.
High resolution spectra show that the Ly$\alpha$ emitting gas has complex 
kinematic structure.
The gas contained within the radio structure has a relatively
high velocity width ($\sim$1500 km s$^{-1}$ FWHM). 
The component of the Ly$\alpha$ emission that coincides with the
bend in the radio structure
is blueshifted with respect to the peak of the emission by 1100 km s$^{-1}$. 
There is low surface
brightness Ly$\alpha$ emission aligned with, 
but extending 40 kpc beyond 
both sides of the radio source.
This halo has a narrow velocity width ($\sim$250 km
s$^{-1}$ FWHM) and a velocity gradient of 450 km s$^{-1}$ over the extent of
the emission.

The
presence of the quiescent Ly$\alpha$ component aligned with the AGN axis, but
outside the radio source, is strong evidence that photoionization by 
anisotropically
emitted radiation from the active nucleus is occurring.
Various mechanisms for the origin and kinematics of the Ly$\alpha$ halo
are discussed. Because the halo extends beyond the radio structure with
less violent and more ordered kinematics than inside the radio structure,
we conclude that the outer halo and its kinematics 
must predate the radio source. 
The ordered motion may be large-scale rotation caused by the accretion
of gas from the environment of the radio galaxy or by a merger.
Although alternatively the halo may be caused by a massive outflow,
we argue that bulk inflow of the emission line gas
is inconsistent with the most likely orientation of the radio source. 

The large velocity-width of the Ly$\alpha$ gas contained
within the radio source compared to that of the
outer halo suggest a direct interaction of the radio source with the gas.
The spatial correlation of enhanced, blue-shifted Ly$\alpha$ emission
and the sharp bend of the radio structure suggest that the emission line
gas could have deflected the radio jet. The impact
of the jet could have accelerated the
gas at this position and may have locally enhanced the Ly$\alpha$ emission.


Extended faint optical continuum emission is aligned
with the principal radio axis, a phenomenon commonly observed in high redshift
radio galaxies. This emission it does not follow the bending of the radio jet,
indicating that, at least in 1243+036,
models invoking scattering of continuum radiation from the AGN 
as the cause of this alignment are favoured.

\end{abstract}

\section{Introduction}
Finding galaxies at large redshifts is important for our understanding of
galaxy formation in the early Universe. Selecting ultra steep spectrum radio
sources ($\alpha < -1$, see Sect.~2) 
has proven to be the most efficient way of finding distant galaxies
(e.g. Chambers et~al. 1990\nocite{cha90}; McCarthy 1993\nocite{mcc93a}).
Searches based on this method have resulted in the discovery of many high 
redshift ($z>2$) radio galaxies (HZRGs), but few with $z>3$ 
(\cite{lil88,cha90,mcc90b,eal93c,rot93}).
In the last few years, we have carried out
a programme to enlarge the sample of HZRGs, 
by selecting ultra steep spectrum (USS) radio sources (e.g. R\"ottgering 
1993)\nocite{rot93}.
This project has included radio imaging (\cite{rot94}), optical broadband
imaging (\cite{rot95}) and spectroscopy of potential distant galaxies. 
The optical observations were carried
out as a Key Programme at the European Southern Observatory in Chile. 
Our USS radio source survey has resulted in the discovery of more than 30
radio galaxies at redshifts $z>2$ (\cite{rot93,oji94a,rot95a}). We have now
started follow up observations to study individual objects in more detail.
This includes high resolution spectroscopy and narrow band imaging of the 
Ly$\alpha$ emitting gas, infrared spectroscopy and a sensitive high resolution
radio survey of the objects.
Here we report on the most distant galaxy discovered in the ESO Key Programme,
1243+036 (= 4C\,03.24) at z=3.57.
This is one of the most distant galaxies reported until now and its redshift
is only exceeded by 4C\,41.17 at $z=3.8$ (\cite{cha90}) and 
8C1435+63 at $z=4.25$ (\cite{lac94}).
\newline
In Sect.~2 we describe the selection of the object,
the observations and data analysis. In Sect.~3
the observational results are presented and physical parameters are deduced.
Subsections describe the determination of the redshift and the kinematics of
the Ly$\alpha$ gas (3.1), the properties of the radio source (3.2), the
morphology of the optical continuum and Ly$\alpha$ emission (3.3) and a
summary of the various radio and optical components (3.4). In Sect.~4 
we discuss the observed characteristics of 1243+036
and in particular the kinematics of the ionized gas.
We discuss the ionization and density of the Ly$\alpha$ gas (4.1), the 
kinematics of the ionized gas and scenarios for the origin of the outer halo
(4.2), the bent radio structure and its relation to the ionized gas (4.3) and
the aligned optical continuum emission (4.4).
We summarize our conclusions in Sect.~5.
\newline
Throughout this paper we assume a Hubble constant of 
$H_0=50$ km s$^{-1}$ Mpc$^{-1}$ and a deceleration parameter of $q_0=0.5$.

\section{Selection and Observations}
As a basis for our survey to find distant galaxies, 
several radio catalogues at different
frequencies and different overlapping regions of the sky were combined to
select samples of USS (radio spectral index $\alpha < -1$, where 
S$_\nu$ $\propto$ $\nu^{\alpha}$, with S$_\nu$ the flux density and $\nu$
the frequency) radio sources 
(\cite{rot94}). The samples were
selected at a range of frequencies from 38 MHz to 408 MHz. The radio galaxy
1243+036 is part of the 178 MHz selected sample (\cite{rot94}), and is 
also known as 4C\,03.24.
\newline
The samples of USS sources were then observed with the VLA as a preliminary
for optical imaging and spectroscopy (\cite{rot94}).
\newline
The $1.5''$ resolution VLA image of 1243+036 from R{\"o}ttgering et~al. 
(1994)\nocite{rot94} at 1465 MHz shows a simple, although bent, double
morphology, with a full angular extent of 8$''$. The source has a total flux
density at 1.5 GHz of 310 mJy, and has a very steep integrated radio spectrum, 
with
a spectral index, $\alpha$, between 178 MHz and 2700 MHz of $-1.3$.
\newline 
1243+036 was optically identified from $R$-band CCD imaging of these samples
of radio sources using the ESO-MPI 2.2m telescope. These observations are
described in R{\"o}ttgering et~al. (1995b) \nocite{rot95}.
\newline
Here we describe low resolution spectroscopic observations to determine the 
redshift of 1243+036 and follow up observations using high resolution 
spectroscopy and imaging of the Ly$\alpha$ gas and the radio source to
study the radio galaxy in more detail.
A summary of the various observations is given in Table 1.

\subsection{Optical Spectroscopy}
\subsubsection{Low resolution}
Low resolution optical spectroscopy was carried out on the ESO New Technology
Telescope (NTT), using the ESO Multi-Mode Instrument (EMMI). The detector was
a Thomson CCD with 1024$^2$ pixels that has a scale along the slit
of $0.37''$ per pixel. The
spectrum covered the wavelength region from 4100 \AA\ to 7900 \AA. A $2''$ wide
slit was used, 
oriented along the radio axis (position angle 152$^{\circ}$, PA1 in Table 1),
giving a spectral resolution of 12 \AA\ (FWHM).

Four separate exposures of 30 minutes each were taken on 21 March 1991. The
seeing during the observations was $\sim 1.5''$. The conditions were
photometric and the spectroscopic standard star EG54 was observed for flux
calibration (\cite{oke74}). The raw spectra were flat-fielded, sky-subtracted,
flux-calibrated and corrected for atmospheric extinction 
using the longslit package in the IRAF reduction package 
of the U.S. National Optical Astronomy Observatory. 
The accuracy of the
wavelength calibration was $\sim$3\AA\ and we estimate that the flux
calibration is good to $\sim$10\%. 
A one-dimensional spectrum
of 1243+036 was extracted using an aperture of $7''$ to include all emission
from the detected spatially extended emission line (see Sect.~3). 

\subsubsection{High resolution}
High resolution optical spectroscopy of the Ly$\alpha$ emission line was carried
out with EMMI on the ESO NTT on 14 April 1994. The detector was a Tektronix CCD
having 2048$^2$ pixels with a scale along the slit of $0.27''$ per pixel. 
The CCD was binned
2x2 giving an effective scale along the slit of $0.54''$ per pixel. 
Using a $2.5''$ wide slit with
ESO grating 6, the spectral resolution was 2.8 \AA. Four 1 hour exposures were
taken with a slit orientation along the radio axis (152$^{\circ}$, PA1), 
and one 1 hour exposure with the
slit orientation angle orthogonal to the radio axis (at 62$^{\circ}$, PA2).
Conditions were photometric and the seeing was approximately $1''$. The
spectra were reduced in the same manner as the low resolution spectra
described above. The skyline at 5577 \AA\ leaves strong residuals
after sky subtraction and this region of the spectrum was therefore not taken
into account in the analysis.

\subsection{Radio}
After the initial VLA ``snapshot'' observations described in R{\"o}ttgering
et~al. (1994) \nocite{rot94}, further radio observations of 1243+036
were made on August 20, 1991 and March 18, 1994, using the VLA in 
`A', configuration (see Table 1). 
The March 18, 1994 observations were part of a
high resolution survey of a large sample of  powerful radio galaxies at
redshifts $z > 2$ (\cite{car95}). All observations were made 
using a 50 MHz bandwidth.

Data processing was performed using the Astronomical Image Processing
System (AIPS). The system gains were calibrated with respect to the standard
sources 3C\,48 and 3C\,286. We estimate that the
flux scale is accurate to better than $1$\% at all frequencies. Phase
calibration was performed using the nearby calibrator 1236+077. The (on-axis)
antenna polarization response terms were determined, and corrected-for, using
multiple scans of the  calibrator 0746+483 over a large range in parallactic
angle. Absolute linear polarization position angles were measured using two
scans of 3C\,286 separated in time by a few hours. From the difference in
solutions between these two we estimate the systematic uncertainty in the
observed polarization position angles to be $\sim$1$^{\circ}$ 
at all frequencies.

The calibrated data were then edited and self-calibrated using standard
procedures to improve image dynamic range. We have also combined the 1.5 GHz
data presented in R{\"o}ttgering et~al. (1994) \nocite{rot94} with the
new data from August 20, 1991. The gridded visibilities were uniformly
weighted to maximize spatial resolution. Images of the three Stokes
polarization parameters, I, Q and U were synthesized, and all images were
CLEANed down to the level of 2.5 times the theoretical RMS on the image, using
the algorithm as implemented in the AIPS task MX. The observations at the 
different frequencies were added in the image plane, weighted by the root rms of
the images, to produce the final maps at 8.3 GHz, 4.7 GHz and 1.5 GHz.

\subsection{Narrow Band Imaging}
Narrow band imaging was carried out with the ESO NTT using the Superb Seeing 
Imager (SUSI). A narrow band filter (ESO filter \# 731) was used that has a 
central wavelength of
5571 \AA\ and a bandpass width of 60 \AA\ (FWHM). The detector was a
Tektronix CCD with 1024$^2$ pixels and a scale of $0.12''$ per pixel. Four
separate exposures of 30 minutes were taken each with a seeing better than 
$0.6''$. Each image was shifted by about $20''$ with respect to the previous
to minimize problems due to flat-fielding and to facilitate cosmic ray removal.
Image reduction was carried out using IRAF and AIPS. The individual images were
bias-subtracted and flat-fielded using twilight exposures. The
images were then registered using shifts determined from the peak positions of
several stars on the CCD near the object and coadded after cosmic ray removal. 
To improve the signal to
noise, the resulting 
image was smoothed with a Gaussian function having a full width 
half maximum of
$0.37''$ (3 pixels) giving a final resolution, as measured from stars on
the final image, of $0.6''$.
Several stars that were visible on both the $R$-band image and the narrow band
image were used to register the two images to an accuracy of $\sim0.1''$.
Absolute astrometric calibration was carried out using the Guide Star Catalogue
(GSC) image processing system of the Space Telescope Science Institute
(\cite{las90}) resulting in an accuracy of $\sim0.7''$.

\subsection{Infrared}
Infrared spectroscopy was performed on June 14, 1994, with the United
Kingdom Infrared Telescope (UKIRT). The [O\,III]
$\lambda$4959,5007 and H$\beta$ $\lambda$4861 lines, redshifted into the
$K$ band window, were observed using the CGS4 spectrograph in the long camera
mode, giving a pixel scale along the slit 
of $1.5''$ and a spectral resolution of 0.006 micron
($\sim800$ km s$^{-1}$).
The object was observed in the beam-switching mode:
the object was switched every 3 minutes 
between two different positions on the slit.
The two series of
observations were subtracted from each other to produce a sky-subtracted
two-dimensional spectrum with a positive and a negative spectrum of the object.
Further details of this observing method are described in Eales \& Rawlings
(1993).\nocite{eal93b}
The spectrum was extracted in a $7.5''$ aperture (similar to that used for 
the optical low resolution spectroscopy) to include all extended line 
emission and wavelength and flux-calibrated 
using the ``longslit'' package as incorporated in IRAF.

A $K$-band image was obtained and reduced for us by Steve Eales and 
Steve Rawlings 
using IRCAM at the United Kingdom Infrared Telescope at Mauna Kea, Hawaii,
in March 1993. It has a pixel scale of $0.62''$x$0.62''$ 
and the total on source 
integration time was 1750 seconds. For more details on the
IRCAM observations and reductions see Eales \& Rawlings (1995).\nocite{eal95}

\section{Results}
\subsection{Low resolution spectroscopy: Determination of the redshift}
The low resolution optical spectrum showed one bright emission line (Fig.~1).
This line is
spatially extended by $7''$, has an observed equivalent width of
1800\AA\ $\pm$300\AA\ and an integrated flux of 2.7$\pm$0.3 x 10$^{-15}$
erg s$^{-1}$ cm$^{-2}$ in the $7''$ aperture. 
No further emission lines were seen in this spectrum. The
faint continuum of 1243+036 is detected redward of the emission line. 
The continuum is not detected blueward of the emission line, giving a lower
limit to the continuum drop of a factor 2.5 $\pm$1. The large equivalent
width of the line and the drop in continuum level at the emission line
(similar to the continuum drop across the Ly$\alpha$ emission line in distant
quasars due to intervening Ly$\alpha$ forest absorption, e.g. Steidel \& 
Sargent, 1987\nocite{ste87}, and references therein) along with the faint
$R$ band identification and the small ultra steep spectrum radio source led us
to tentatively identify the line with Lyman $\alpha$ at $z=3.57 \pm 0.01$. The
redshifted Ly$\alpha$ line lies close to the strong 5577 \AA\ skyline making
a more exact redshift determination difficult from the low resolution
spectrum.

The infrared spectroscopic observation showed two bright emission lines in 
$K$ band (Fig.~2). 
These were identified with the [O\,III]$\lambda\lambda$4959,5007\AA\ 
emission lines, indicating that the supposed 
redshift is indeed correct. There is also a hint of H$\beta$ 
$\lambda$4861\AA\ emission.
The [O\,III]5007 has a flux of 3$\pm$0.5 x 10$^{-15}$ erg s$^{-1}$ cm$^{-2}$,
comparable to that of Ly$\alpha$, and has a spatial extent of at least
$5''$, i.e. comparable to the extent of Ly$\alpha$ in the low resolution 
optical spectrum. The spectrum of the emission lines in the infrared has
much lower signal to noise ratio than that of Ly$\alpha$ in the optical. 
Therefore any faint [O\,III] emission in the region outside the radio lobes,
if the [O\,III] would be equally extended as the Ly$\alpha$ (see below),
could not be detected.

\subsection {Radio Polarimetric Imaging}

Figure 3a shows the total intensity image at 8.3 GHz with
0.23$''$ resolution. Figure 3b
shows the image at 4.7 GHz 
with a resolution of 0.43$''$ and
Fig.~4 shows an image at 1.5 GHz with $1''$ resolution,
produced from the combined data 
from R{\"o}ttgering et~al. (1994)\nocite{rot94}, 
and later observations at the same frequency. 
A number of features have been labeled for reference.
Table 2 lists the positions of the peaks,
the peak surface brightnesses (I) and integrated fluxes (S, 
measured in a rectangular box
around the components) for the various source components.

The detailed source structure is best represented in the 8.3 GHz image.
The northern lobe (designated `A') has a compact component with a 
`tail' extending to the north. The southern lobe (designated `B') 
again has a compact component (B1) close to the nucleus, and  a linear series 
of knots (the southern `jet'),
extending towards the south from B1, with a slight extension
to the west at its southern end.
In between these two lobes is a faint region of emission (designated
`N'),  that has a core--jet structure with a length of about 0.8$''$,
oriented at a position angle
of $-24^{\circ}$, i.e. roughly aligned with the axis defined by components
A1, N, and B1. The southern jet has an angle of 30$^{\circ}$ away from this
axis.

The spectral index image of the 8.3 GHz and 4.7 GHz maps convolved to 0.43$''$
resolution (with a cut-off where the signal to noise ratio in the original maps 
is smaller than 8) is shown in Fig.~5a, and of the 4.7 GHz and 1.46 GHz
maps convolved to $1''$ resolution is shown in Fig.~5b. 
Spectral index values between 8.3 GHz and 4.7, GHz determined from the surface
brightness of the various source components convolved to the same resolution,
are also listed in Table 2. Most components of the radio source
have steep spectra, with indices typically $<-1.4$. The two exceptions are
components B1 and N, both of which have flatter spectral indices of $\sim
-1.0$. One curious aspect of the 1243+036  radio structure is that the
spectral index distribution steepens with distance from  the source centre in
both components A and B. This is unlike most high power radio galaxies, which
show the flattest spectral indices at the radio hot spots situated at the
extremities of the radio source (\cite{car91,ale87}).
This spectral steepening may be due to 
synchrotron radiation losses of the highest energy electrons.

We assume minimum energy conditions
(see Miley 1980; \nocite{mil80b} Pacholczyk 1970\nocite{pac70})
to derive a minimum magnetic 
field strength. In calculating the minimum energies we assume that the radio 
source is cylindrically symmetric, the plasma has unit filling factor and that
the synchrotron spectrum extends from $10^7$ to $10^{11}$ GHz.
For the components with high frequency spectral index $\alpha^{4.7}_{8.3}<-1.0$
a two--slope power-law spectrum was assumed with a spectral index below 4.7 GHz
(observed frame) of $-1.3$, equal to the integrated source spectral index
between 178 MHz and 2700 MHz. 
The derived parameters for the various components are in Table 3.
\newline
We can estimate the density of the external medium
assuming
that the minimum pressures are balanced by either ram pressure, due to the
propagation of the radio plasma at a velocity of a few thousand km s$^{-1}$
(e.g. Cygnus A, Carilli et~al. 1991\nocite{car91} and references 
therein), or by thermal
pressure from a hot ($\sim10^7$ K) external medium, which is often found
to be present around powerful radio galaxies and in clusters at low redshifts
(see Crawford et~al. 1988; Baum et~al. 1989\nocite{cra88,bau89b}
and references therein).
For the typical source parameters, we find that the external
density ranges from 0.1 cm$^{-1}$ (ram pressure) to several tens cm$^{-1}$
(thermal pressure), see Table 4.

Assuming that the spectral steepening of the outer components relative to
the inner components is due to synchrotron radiation losses, 
we crudely estimate the synchrotron age of the 
emitting radio plasma in a few components from the 
break frequency where the radio spectrum steepens and the magnetic field
strength.
We estimate the restframe break frequency by fitting a ``Kardashev-Pacholczyk
model'' to the spectral 
index distribution in Figs. 5a and 5b and assuming
that the input spectral index $\alpha_{in} = -0.7$ 
(spectral index at frequencies below the synchrotron break). 
For more details about spectral ageing modelling
see Carilli et~al. (1991) \nocite{car91}. 
The synchrotron age of the radio plasma is given by:
$t_s = 1610B^{-3/2}\nu_B^{-1/2}$ Myr with B in $\mu$G and $\nu_B$
in GHz, see Carilli et~al. (1991).\nocite{car91}
We note however that there are 
large uncertainties in these ages, because of the strong dependence on the
magnetic fields that are
derived from the high restframe frequencies only. 
In any case, the derived synchrotron ages give an impression of the age
of the outer components (B4) relative to the inner components (B1).
The derived synchrotron ages for the components A1, B1 and B4 are listed in 
Table 3.

Images of polarized intensity from 1243+036 are shown in Fig.~6,
along with vectors representing the position angle of the 
electric field vectors. The only component that shows
substantial polarization at both frequencies 
is the southern jet (components B2, B3, and B4). 
Component A shows weak polarization at 8.3 GHz, and no polarized
emission at 4.7 GHz, and components B1 and N show no
polarized emission at either frequency. The fractional
polarization of these various components (4$\sigma$ upper limit for B1) 
are listed in Table 2. Also listed are the depolarization ratios, 
D$_{8.3}^{4.7}$, for the components with detected polarized signal,
where D$_{8.3}^{4.7}$ is the ratio of fractional polarization at
4.7 GHz to that at 8.3 GHz. Component B4 shows no significant depolarization
between 4.7 GHz and 8.3 GHz, while component B2 has a 
value of D$_{8.3}^{4.7}$ = 0.55, and A1 only has an upper limit (4$\sigma$)
of D$_{8.3}^{4.7}$ $<$ 0.3.

We estimate that the uncertainties in observed position angle
in the  brighter polarized regions are about 2$^{\circ}$, based on signal-to-noise 
and possible systematic errors of $\sim$1$^{\circ}$ discussed
in Sect.~2. Overall, we feel that a position angle change of $\ge$ 6$^o$
between 8.3 GHz and 4.7 GHz could have been measured with this data at the
peaks in polarized intensity. No such change is seen, implying
an upper limit to the rest frame rotation measures of $\sim$1000
rad m$^{-2}$. This limit applies to the regions that are not depolarized.

The projected magnetic field distribution is shown in Fig.~7
overlayed on the 8.3 GHz total intensity map.
In the southern region, the projected field is parallel to the jet, 
and then bends towards the
west in the vicinity of knot B4. In the north the  
projected field is parallel to the `tail' of component A towards the northwest. 
Towards the west of knot B4 there is slightly extended radio emission
visible at all three radio frequencies.
This emission and the change of the magnetic field direction at that position
suggest that this may be outflow from the final hotspot (B4) into the radio
lobe.

\subsection{Optical and IR Imaging}

\subsubsection{Broad band imaging}

The $R$ band image identifying the radio source is from R{\"o}ttgering 
et~al. (1995b).\nocite{rot95}
The optical
identification is a faint $2''$--$3''$ extended galaxy with an integrated
magnitude of $R=22.5$ (\cite{rot95}). 
In the low resolution optical spectrum, continuum emission is detected but
no emission lines were detected in the
$R$ band region of the spectrum. The strongest emission
line that might affect the $R$ band flux is CIV$\lambda$1549\AA\ 
(expected at 7079 \AA) but was not
detected, so that the $R$ band does not contain any
appreciable contribution (less than 1\%) from line emission. 
The $R$ band image therefore
represents the continuum emission from 1243+036.
The optical galaxy is clearly aligned with the main radio
axis, as is common in high redshift radio galaxies (\cite{cha87,mcc87a}). 
In Fig.~8 the same $R$ band image is rebinned and 
smoothed with a Gaussian of
FWHM $1.0''$ and overlayed with the contours of the 8 GHz VLA map. There
appears to be faint extended continuum emission in the direction of the
principal radio axis that extends
beyond the radio lobes (indicated A and B in Fig.~8).
Although at a low surface brightness, the total extended
emission is detected at a signal to noise of $\sim$10 in both
north (A) and south (B).
Three more small blobs at the 10$\sigma$ significance level are
seen on the smoothed frame (indicated C, D and E) in the
region to the southwest of 1243+036, while elsewhere on the CCD frame 
such a concentration of emission features is not observed.

The $K$-band image obtained at UKIRT gave 
a magnitude for 1243+036 in a $5''$ aperture of $K$=18.8
$\pm$ 0.2. However, the $K$-band flux includes the strong redshifted 
[O\,III] emission line (see Fig.~2). 
After correcting for the [O\,III] flux as determined from the infrared 
spectroscopy,
we find a continuum $K$ magnitude of 19.3 $\pm$ 0.3. This $K$-band
magnitude is typical for radio galaxies at these redshifts (e.g. Eales \& 
Rawlings 1993; Eales et~al. 1993b).\nocite{eal93b,eal93a}
The signal to noise ratio of the detection was low and the image is therefore 
not shown.
A Keck-telescope image in $K$ band shows
that it is aligned with the optical and radio structure (van Breugel, private 
communication).

\subsubsection{Narrow band Lyman $\alpha$ imaging}
A contour image of the Lyman $\alpha$ emission from 1243+036 is
shown in Fig.~9. The peak, measured as the position of the maximum of
a twodimensional parabolic fit to the brightness distribution, is at 
12$^h$ 45$^m$ 38.36$^s$, 03$^o$ 23$'$ 21.1$''$.
The Ly$\alpha$ emission has a complex morphology:
(1) `Cone-shaped' emission
extends along the radio axis towards the south--east for about 2$''$.  
(2) On a scale of $\sim5''$ the gas shows two `arms' of emission both 
extending almost due south, originating
from the southern end of the inner cone of emission. 
(3) To the north there is faint emission which extends about 3$''$
from the Ly$\alpha$ peak at a position angle of about $-25^{\circ}$.
(4) The lowest surface brightness Ly$\alpha$ emission on scales of $8''$
has a clumpy or filamentary structure. This clumpy structure 
is at a level of 4--6$\sigma$.
(5) There is a small region of enhanced line emission at the southern end 
of the 
Ly$\alpha$ emission, due south of the `gap' between the two `arms', mentioned
above.

\subsection{High resolution spectroscopy}

The positions of the slits used in the 2.8 \AA\ resolution spectroscopy are 
shown in Fig.~14.
The spectrum with the slit along the radio 
axis (PA1=152$^{\circ}$) shows the velocity field of the central and most
spatially extended Ly$\alpha$.
The overall shape of the central velocity profile (Fig.~10)
has a width of $\sim 1550\pm 75$ km s$^{-1}$ (FWHM). The line does however
not have a smooth shape. It has several small dips (or peaks), 
especially in the blue wing of
the Ly$\alpha$ emission line. These small dips at 5533, 5538, 5544 and 
5550 \AA\ were present in the individual exposures.

The two-dimensional Ly$\alpha$ spectrum in PA1, smoothed with a Gaussian of 
FWHM 2 pixels ($1.1''$) to enhance fainter extended Ly$\alpha$ emission,
is shown in Fig.~11. 
Most of the flux is in the centre and has a spatial
extent of $\sim 2''$ that peaks at 5557 \AA\, i.e. $z=3.5699\pm0.0003$.
The total flux of Ly$\alpha$ is better determined from
the high resolution spectrum, since the subtraction of the skyline at 
5577 \AA\ can be carried out with a much greater accuracy than in the low 
resolution spectrum. 
The integrated
Ly$\alpha$ flux in an aperture of $11''$
is $2.9 \pm 0.2$ x 10$^{-15}$ erg s$^{-1}$ cm$^{-2}$ 
corresponding to a luminosity of $\sim2.7$ x 10$^{44}$ erg s$^{-1}$.
In a larger aperture of $24''$, including the most extended emission
seen in the smoothed two-dimensional spectrum (see below and Fig.~11), 
the Ly$\alpha$ flux is 
$3.2$ x 10$^{-15}$ erg s$^{-1}$ cm$^{-2}$.

A separate velocity component in Ly$\alpha$ emission is seen at about 1100 
km s$^{-1}$ blueward and $2''$ southward along the slit from the central 
Ly$\alpha$ peak.
This component contains almost 10\% of the total Ly$\alpha$ 
flux and has a velocity width of $\sim 1200\pm$100 
km s$^{-1}$ (FWHM).
Its position coincides with the brighter of the two `arms' of Ly$\alpha$
emission seen in the Ly$\alpha$ narrow band image and with radio component B1
(see below).
\newline
There is also a remarkable, extended narrow component ($\sim250$ km s$^{-1}$ FWHM)
that clearly shows a velocity shear on both sides of the nucleus. It has a 
total spatial extent of $\sim 20''$, corresponding to 136 kpc,
much larger than the size of the radio source ($8''$).
The velocity difference between the northern and southern extent of this
emission is $\sim 450$ km s$^{-1}$.
\newline
Note that this very extended Ly$\alpha$ component was not seen in the narrow
band Ly$\alpha$ image.
This is because the resolution of the spectrum is very well
matched to the velocity width of the most extended Ly$\alpha$ gas, 
whereas the narrow band filter has a much larger spectral width.
Also the presence of the 5577\AA\ skyline adds a 
considerable amount of noise in the narrow band image.
\newline The small dips seen in the one-dimensional spectrum are also clearly
visible on the two-dimensional spectrum (Fig.~11). Especially 
noteworthy is the feature at $\sim5533$\AA , that extends spatially from the
central Ly$\alpha$ emission to the separate velocity component at B1.

The high resolution spectrum taken with the slit perpendicular to the radio
axis (PA2=62$^{\circ}$, Figs. 12 and 13)
shows only the Ly$\alpha$ emission from the brightest inner part as 
seen on the Ly$\alpha$ narrow band image. The emission is spatially resolved. 
The emission has the same overall velocity width
and peaks at the same wavelength as the spectrum taken along the radio axis. 
The two-dimensional spectrum shows a depression in the
Ly$\alpha$ emission blueward of the peak in the central rows of the 
spectrum over about 6\AA\ ($\sim300$ km s$^{-1}$). 
The Ly$\alpha$ emission blueward of the depression has the same velocity 
as that of the separate velocity component described above along PA1.
Redward of the depression the Ly$\alpha$ has the same velocity as that
of the brightest inner Ly$\alpha$ in the PA1 spectrum. The signal of the main
Ly$\alpha$ emission is not as strong as in the PA1 spectrum. 
Therefore it seems that
the $2.5''$ wide slit was not exactly positioned at the peak of Ly$\alpha$
and the blueshifted Ly$\alpha$ emission at the position of B1 is also 
at least partly included (see Fig.~14). 
\newline
The very extended low surface brightness Ly$\alpha$ seen in the spectrum along
the radio axis (PA1) is not seen in the spectrum with the perpendicular slit 
position (PA2).
Although the spectrum at PA2 is not as deep as that at PA1 
(the noise is a factor 2 higher), such faint emission
would have been detected if it was of the same extent and surface brightness
as the emission as observed along the radio axis.

\subsection{Description of the various components}
Figure 14 shows the Lyman $\alpha$ image
of 1243+036 overlayed with the radio continuum emission at 8.3 GHz. 
Figure 8
shows the same radio image overlayed with the $R$ band optical image.
Standard VLA astrometry is
accurate to at least 0.2$''$, the optical astrometry has an 
accuracy of $0.7''$ (\cite{rot95}).

The spatial extent of the Ly$\alpha$
emission visible in the narrow band image 
is similar to that of the radio source ($\sim$8$''$).
Component N of the radio source is located at the peak of the line emission
and $R$-band emission.  Since N is also the
flattest spectrum radio component, we identify it as
the centre of the nuclear activity of 1243+036. 
The spatial extent of this radio core--jet 
component (length $\approx$
0.8$''$) is comparable to the spatial extent 
of the highest surface brightness line emission.

The radio galaxy 1243+036 exhibits two trends seen in samples of
high redshift radio galaxies ({\sl cf.} McCarthy 1993a)\nocite{mcc93a}. 
Firstly, the alignment between the major axis
of the line emitting gas and the optical continuum 
with the principal axis of the radio source. 
Secondly, the emission line gas is distributed
asymmetrically with respect to the nucleus. 
In most powerful radio galaxies the
emission line gas is brightest on the side of the closest radio lobe,
suggesting interaction between the radio source and the ambient gaseous
medium (\cite{mcc91a}). 
Regarding the entire extent of the radio lobes of 1243+036, 
the southern radio lobe (B) is not the closest to the nucleus (N). However, 
regarding
the brightest hotspots in the lobes, A1 and B1, B1 is the
closer with an armlength ratio of Q=1.2
Further, the southern lobe is the most distorted and 
brightest, indicating a stronger interaction with the environment.

In addition to these two trends, there are three further
correlations in the radio and Ly$\alpha$ properties.
Firstly, both
the line emitting gas and the radio source show peculiar structure
at the position of component B1. The radio source bends by about 
30$^{\circ}$ towards the south.
The line gas has a local peak at this position, and 
the long slit spectra show that 
the line emitting gas in this region is blueshifted by $\sim1100$ km 
s$^{-1}$ with respect to the central ionized gas.
The southern jet follows closely
along the brighter of the two southern `arms' of line emission.
Secondly, the emission line gas is
seen to encompass the northern radio source component A.
The high resolution spectrum (Fig.~11)
shows that the Ly$\alpha$ has a low surface brightness component that
extends far beyond the radio lobes.
Thirdly, the emission line gas inside the radio structure is brighter and
has a much larger velocity dispersion than the emission line gas outside
the radio structure.

Summarizing the results from the high resolution spectroscopy and narrow band
imaging of the Ly$\alpha$
gas we can distinguish between three separate components of the emission line
gas: 

\noindent (i) The high velocity dispersion gas inside the radio structure, 
the {\it ``Inner halo''}. 
\newline
The Ly$\alpha$ emission within a radius defined by the maximum extent of the 
radio structure is dominated by the bright central emission which has
a shape of a cone with a large velocity dispersion ($\sim1550$ km s$^{-1}$ 
FWHM).
Such a velocity dispersion is typical for the
extended emission line gas in high redshift radio galaxies 
(\cite{mcc93a,mcc89a}).

\noindent (ii) Enhanced Ly$\alpha$ emission 
with a blueshifted velocity at the position of radio knot B1.
\newline
The region of
enhanced Ly$\alpha$ emission at the position of radio knot B1, in the brighter
of the two southern arms, has a clear velocity difference ($\sim1100$ km
s$^{-1}$) relative to the central Ly$\alpha$. Its velocity dispersion
($\sim1200$ km s$^{-1}$ FWHM) is similar to the rest of the Ly$\alpha$ gas 
in the inner halo.

\noindent (iii) Ly$\alpha$ emission extended far beyond the radio source,
the {\it ``Outer halo''}.
\newline
The very extended Ly$\alpha$ emission with a velocity shear of $\sim450$ km 
s$^{-1}$ has a total spatial extent of $\sim20''$ (136 kpc), 
much larger than the radio source ($\sim8''$). 
This indicates that this component of the gas has 
no direct relation with the radio plasma. 
This
extended Ly$\alpha$ was seen with the slit of the spectrograph along the
principal radio axis. The spectrum perpendicular to the radio axis showed
that in this direction there is no Ly$\alpha$ emission of such extent.

\section{Discussion}

\subsection{Ionization, density and confinement of the Ly$\alpha$ emitting 
clouds and density of the external medium}
Several mechanisms have been considered for the ionization of the extended 
emission line clouds in powerful radio galaxies. These include ionization
by UV radiation from hot stars, shock ionization from the interaction with 
the radio lobes and photoionization by the active nucleus. The equivalent
width of the Ly$\alpha$ emission in most high redshift radio galaxies is several
times larger than can be produced by hot stars (\cite{mcc93a,chr93}).
In general, the line ratios in radio galaxies are not well reproduced by
shock ionization nor ionization by hot stars
and indicate that photoionization by nuclear UV radiation appears to be
the dominant mechanism of ionization (\cite{bau89b,bau92,mcc93a,fer86,fer87}).
Anisotropic photoionization by an obscured nucleus has been argued to explain
the deficit of observed UV photons from radio galaxies and Seyfert-2 galaxies
relative to the amount needed to produce the emission line luminosities
(\cite{ant85}). This concept has been an important constituent
of orientation-based unification models of 
radio galaxies and quasars were developed, which can explain many observed
correlations and anomalies (\cite{sch87b,bar89,hes93}).
In these models radio-loud quasars and radio 
galaxies are intrinsically similar objects but observed at different angles.
The radiation from
the nucleus escapes preferentially within a cone along the 
radio axis and ionizes the gas clouds in its path.
In quasars we can see inside the ionizing cone
and see the broad-line emission region close to the active nucleus, while
in radio galaxies the nucleus is obscured so that 
we cannot see the ionizing radiation directly, but see the
extended (narrow) emission line region that is ionized by the nuclear continuum.
The alignment of the ionized gas with the radio axis in most powerful radio
galaxies argues in favour of these models, especially the presence of ionized
gas outside the radio lobes. The Ly$\alpha$ emission in the outer halo in
1243+036, outside the radio lobes, but aligned with the principal radio axis,
rules out shock ionization by the radio source and strongly argues for
photoionization by anisotropic nuclear emission.
4C\,41.17 at z=3.8 (\cite{cha90})
and 6C\,1232+39 at z=3.2 (\cite{eal93c}) are other examples of high redshift
radio galaxies with Ly$\alpha$ emission extending beyond the extent of the radio
emission. 
The ``cone'' shaped morphology of the innermost $2''$ of 
the Ly$\alpha$ emission of 1243+036 would not be expected from shock ionization
and further supports the idea that the Ly$\alpha$
halo of 1243+036 is indeed photoionized by anisotropic nuclear
continuum radiation. The shape of this innermost Ly$\alpha$ emission region 
indicates an opening angle for the cone of nuclear 
ionizing radiation of between 35$^{\circ}$ and 50$^{\circ}$.

Although we have very little information on the physical conditions of the
ionized gas we can estimate the density and mass of the emission line gas,
assuming photoionization, from
the flux and extent of the Ly$\alpha$ emission along the lines of McCarthy
et~al. (1990) \nocite{mcc90a} for 3C\,294 and Chambers et~al.
(1990) \nocite{cha90} for 4C\,41.17. We shall use three different approaches to
estimate the density.

I. Following McCarthy (1990)\nocite{mcc90a}, the Ly$\alpha$ luminosity 
is related to the density of the ionized gas, assuming case B 
recombination:
$L=4$x$10^{-24}n_e^2 f_v V$ ergs s$^{-1}$, where $V$ is the total volume 
occupied
by the emission line gas and $f_v$ the volume filling factor of the gas. From
this then follows the mass in ionized gas M$\approx n_e m_{proton} f_v V$.
Because we have no means of measuring the filling factor and density 
simultaneously in 1243+036, we have to estimate the volume filling factor.
From direct measurements, using sulphur lines, 
of the density of line emitting gas in low redshift
radio galaxies, filling factors in the range 10$^{-4}$--10$^{-6}$ and 
typically of order $\sim10^{-5}$ have been found (\cite{bre85d,hec82}).
Assuming a filling factor $f_v\sim10^{-5}$ we give in Table 5
the density of several components of the Ly$\alpha$ halo and their inferred 
masses. The sizes of 
components were measured from the Ly$\alpha$ image and high resolution 
spectroscopy and were assumed to be cylindrically symmetric.
The values derived in Table 5 indicate that the density
of the ionized gas falls off with radius as $r^{-2}$.
This would mean that in the case of anisotropic nuclear photoionization the
ionization parameter would be about the same at all radii. This is consistent
with the findings of Baum et~al. (1992)\nocite{bau92} for the
emission line nebulae in powerful radio galaxies at low redshifts and also 
found for the extended line emission around quasars at $z>0.5$ 
(see Bremer et~al. 1992a\nocite{brem92a} and references therein).
The Ly$\alpha$ peak at B1 seems to be an exception. However, as we will
argue in Sect.~4.3, the jet--gas interaction that bends the radio jet at the
position B1, may have shredded dense neutral cores of the emission line 
clouds, therefore increasing the amount of gas exposed to the
nuclear ionizing radiation and increasing the volume filling factor of 
ionized gas, consequently enhancing the Ly$\alpha$ emission
at that position (see Bremer et~al. 1995). 
If the volume filling factor is locally increased 
to 10$^{-3}$, the density decreases to $\sim$100 cm$^{-3}$, 
similar to that in the 
rest of the inner halo. We note that this would infer a higher mass 
($\sim10^8$ M$_{\odot}$) of ionized gas at B1.

II. To prevent the emission line clouds from dispersing on short time scales, they
must be confined by external pressure, as was pointed out by Fabian 
et~al. (1987).\nocite{fab87} We can estimate the density of the
emission line filaments, assuming that they are confined by a hot ($\sim 10^7$
K) external medium, with density derived from the radio data. The emission
line clouds may also be confined by the pressure from the radio lobes, however
the radio lobes in turn must be confined by an external medium, leading
to the same pressure requirement. Further evidence for the presence
of a hot halo comes from the strong depolarization of the radio emission,
that is most likely caused by a hot (X-ray) halo as magnetoionic medium 
(\cite{lai88,gar88}). Other HZRGs (e.g. 4C\,41.17 and B2\,0902+34, 
Carilli et~al. 1994\nocite{car94}) 
show evidence for the presence of a hot surrounding halo.
Unfortunately, at the high restframe 
frequencies observed, the radio emission from the extended lobes is not 
detected and the emission is dominated by the emission from the hotspots.
The hotspots are most likely confined by
the ram pressure caused by their propagation through the external medium and
not by static thermal pressure.
Because of its flatter radio spectrum hot spot B1 appears to be the most
``active'' and we therefore assume the balance between minimum pressure and
ram pressure from B1 to be the best representative pressure. 
Ram pressure confinement
of the advancing hotspot requires a density of $\sim$0.1 cm$^{-3}$ (see Table
4 and Sect.~4.3). Hotspot A1 requires a similar density.
Assuming pressure balance between the hot halo gas ($nT\sim
10^6$ K cm$^{-3}$) and the emission line clouds 
(with a temperature of $\sim10^4$ K)
gives an average density of the ionized gas of $\sim$100 cm$^{-3}$, roughly
consistent with the emission line gas density derived assuming 
$f_v\sim10^{-5}$. 

III. Assuming that every ionized hydrogen atom eventually results in the emission
of a Ly$\alpha$ photon, the Ly$\alpha$ luminosity indicates the necessary 
flux of ionizing photons, Q$_{ion}\approx2$x10$^{55}$ photons s$^{-1}$.
Using this and an estimate of 
the ionization parameter, U, of the gas gives a third estimate
of the density: U=Q$_{ion}/4\pi r^2 n_e c$.
In 1243+036 we have no emission lines from which we could directly estimate
the ionization parameter. However, from photoionization modelling of high 
and low redshift radio galaxies, the
best fit to the emission line ratios is achieved for log U=$-2.5$ 
(\cite{mcc93a,mcc90a}). A similar number is obtained in 3C quasars where Q is 
known (\cite{cra89,cra88}). Thus, assuming log U=$-2.5$ at a characteristic 
radius in the inner halo of $\sim$25 kpc gives a density $n_e\sim$ 3 cm$^{-3}$.
However this assumes the ionizing radiation is emitted isotropically.
If photoionization is anisotropic within a cone of opening angle
$\sim$45$^{\circ}$,
the ionizing flux and corresponding densities are an order of magnitude higher.

We note that the derived densities are only crude estimates due to the
many uncertainties (e.g. filling factor and minimum pressures) and can
therefore be only order of magnitude estimates.

\subsection{Morphology and kinematics of the Lyman $\alpha$ halo}
We shall discuss the kinematics of the emission line gas for each of the 
components as defined in Sect.~3.5 separately.

\noindent (i) {\it Inner halo}
\newline
If the gas motion of the inner halo is
gravitational in origin, the velocity dispersion would imply a virial 
mass of $\sim10^{12}$ M$_{\odot}$ 
inside a radius of $\sim$30 kpc. 
However, the spatial correlation of the high velocity
gas of the inner halo and the radio structure, the low velocity dispersion of
the outer halo and the enhanced and accelerated Ly$\alpha$ gas at radio knot B1,
suggest that the high velocity dispersion of the inner halo gas is
the result of hydrodynamical interactions with the radio plasma.
The interaction of emission line gas with the radio source 
has been suggested earlier as the cause for the high velocity dispersions of 
the extended emission line gas 
in high and intermediate redshift radio galaxies and radio loud quasars and 
there is evidence that
radio sources can entrain and accelerate gas 
(\cite{you81,bre85d,you86,mcc90a,cha90,brem95}).
Calculations by De Young (1986) \nocite{you86} showed that gas entrainment rate
can be up to a few hundred solar masses per year.
\newline
The narrow dips, observed mainly on the blue wing of the Ly$\alpha$ profile of 
the inner halo and at least one extending spatially from the central part to 
the separate 
velocity component at B1, are interesting features that deserve some attention.
Although we cannot exclude that these features are due to true velocity
structure of the line emitting gas, the spatial continuity of one of these 
dips from the central
Ly$\alpha$ to the blueshifted Ly$\alpha$ emission component at B1 
indicates that they are
probably due to absorption by associated neutral hydrogen clouds. 
This also indicates that these absorption clouds must be
in front of the Ly$\alpha$ emission region and have a minimal spatial extent
of $\sim20$ kpc.
\newline
If the emission line gas is ionized by anisotropic nuclear 
photoionization as in the unification models of quasars and radio galaxies,
the absorption systems seen against the Ly$\alpha$ emission in 1243+036
may be clouds similar to the Ly$\alpha$ emitting
clouds, situated outside the cone of ionizing radiation, but in our
line of sight to the Ly$\alpha$ gas (\cite{oji95b,rot95a}).

\noindent (ii) {\it Enhanced Ly$\alpha$ at B1}
\newline
The positional coincidence of the separate velocity component of the emission
line gas and the bend in the radio structure at B1 
suggests a direct interaction of the radio jet and the Ly$\alpha$ gas 
accelerating the gas to its obserrved velocity. The interaction of 
the radio jet with the gas could have locally enhanced the Ly$\alpha$ emission.
This will be further discussed in Sect.~4.3.

\noindent (iii) {\it Outer halo}
\newline
The narrow velocity width (250 km s$^{-1}$ FWHM) of the emission line gas in the
outer halo
compared to that of the Ly$\alpha$ emission in the inner halo and its extent 
to far beyond the radio source
supports the suggestion that this very extended component is not
affected by the interactions that have stirred up the inner Ly$\alpha$ gas,
as mentioned above. 
\newline
The velocity shear observed in the outer halo of $\sim 450$ km s$^{-1}$
may be due to rotation or a flow of gas, i.e. infall or outflow. 
We will discuss these possibilities
and their possible connections and implications for the formation of
powerful radio galaxies in the early Universe.

{\sl a) Rotation}
\newline
The velocity shear might be due to a large scale
rotation of a gas disk in the halo. A gravitational origin of the rotation
of such a large disk implies a mass of $\sim10^{12}$ sin$^{-2}(i)$ 
M$_{\odot}$, where $i$ is the 
inclination angle of the disk with respect to the plane of the sky. 
Baum et~al. (1992) \nocite{bau92} propose that at 
low redshifts rotation of the extended emission line gas in powerful radio
galaxies indicates the presence of dynamically young disks of gas
acquired in a recent interaction or merger with a gas-rich galaxy. 
These rotating emission line regions are, 
however, only on scales of $\sim$20 kpc, while in 1243+036 the outer halo 
is almost an order of magnitude larger.  
Large gaseous halos are common around HZRGs (\cite{mcc93a}) and may imply that
all primeval galaxies in the direct (cluster) 
environment of HZRGs also possessed such large halos.
An interaction between two (primeval) galaxies with large gaseous 
halos may leave kinematical signs of the event 
over a much larger scale than the relatively small rotating 
emission line nebulae that
exist around radio galaxies at low redshifts that according to Baum 
et~al. (1992)\nocite{bau92} are due to interactions. 
Thus, it is possible that
the rotating outer halo of 1243+036 originates from an interaction with a
neighbouring primeval galaxy.
\newline
Alternatively and perhaps more likely, 
the outer halo Ly$\alpha$ emitting gas in 1243+036 may be
a relic of the gaseous halo from which the galaxy has been forming since 
before the radio source switched on.
A rotating disk may have formed during the accretion of gas from the
environment. Numerical simulations by Evrard et~al. (1994)\nocite{evr94} 
of the formation of galaxies in hierarchical clustering scenarios indicate
that rotating disks with radii of several tens of kiloparsecs
naturally form during the formation of galaxies at high
redshift ($z=9$--1).
\newline
Note that, although the gas disk does not have to be rotating exactly 
along the radio axis, it must be within the cone of ionizing radiation from 
the AGN (if it is anisotropically ionized) and, more importantly,
it must be within the slit of the spectrograph that was oriented along the 
radio axis.
The slit has a width of 17 kpc at the redshift of 1243+036. Assuming that a
rotating gaseous disk would not be wider than 35 kpc (one quarter of its
observed diameter) at a radius of $10''$ from the nucleus, this limits the
difference between the orientation of the disk and the radio axis to less than
20$^{\circ}$.
\newline
The fact that we do not see evidence for a
disk in absorption against the bright central Ly$\alpha$ emission does not
argue against the explanation of the velocity shear as originating from a
rotating gas disk. If the observed Ly$\alpha$ gas in the outer halo is ionized
by anisotropic nuclear radiation, we would only expect to see the rotating gas
disk in absorption against the bright central Ly$\alpha$ if we saw the disk
edge on. The size (in the direction of PA2) 
of the bright central Ly$\alpha$ emission region
($\sim5$ kpc) implies an angle between the orientation of the disk and our
line of sight of at least 4$^{\circ}$. 
\newline
It may seem surprising that in the case of 1243+036 the rotation axis of the gas
disk is perpendicular to the radio source axis, because it is generally believed
that radio galaxies have a central black hole fuelled by 
an accretion disk in the inner parts of the galaxy that has its rotation axis
along the black hole's spin axis and the radio axis. Thus, the
the rotation of the outer halo in 1243+036 cannot be a direct extension of 
the inner accretion disk.
However, West (1994)\nocite{wes94} notes that there are sometimes 
remarkable alignments between radio axes of relatively close pairs of
powerful radio galaxies,
and also between the major axes of neighbouring clusters and their central
galaxies. He suggests that this may be a signature of structure formation in
the universe from the largest scale structures down to the scales of the most
massive galaxies in clusters. 
In this case the preferred axis for structure formation is the line between
neighbouring systems.
R{\"o}ttgering et~al. (1996) \nocite{rot95b} find an excess of galaxies 
along the radio axis of powerful radio galaxies.
The orientation of the outer Ly$\alpha$ halo in 1243+036
along the radio axis, would then fit into this scenario of a connection between
the (radio) axes of massive galaxies and large scale structure formation.

{\sl b) Outflow}
\newline
In the case of outflow, it is unlikely that the gas would advance through
the extragalactic medium faster than the radio lobes. Therefore,
the outflow of gas would need to have existed long before the observed
radio source switched on. 
With the maximum velocity of the most extended emission ($\sim250$ km s$^{-1}$)
it would take $\sim3$x$10^8$ years to transport the gas out to a radius of 70 
kpc, an order of magnitude longer than the most plausible age of the radio 
source ($\sim10^7$ years, based on the projected radio size and a radio lobe 
propagation speed of $\sim3000$ km s$^{-1}$, see Carilli et~al. 
1991\nocite{car91} and references therein).
However, we may simply see the galaxy during its most recent period of
activity and it may have developed an extended radio source several times 
during previous active periods in the time needed for the outflow to 
build up the outer Ly$\alpha$ halo. In that case the outer halo may be gas 
swept up and deposited by expanding radio lobes during earlier active periods
of the galaxy. 
\newline
Possibly, 1243+036 may have had an active nucleus before its radio 
activity, ejecting gas along the central black hole's rotation axis.
There is evidence 
for bulk mass outflow of ionized gas close to the nucleus of quasars
at velocities up to 1000 km s$^{-1}$ (\cite{car91a}). In broad absorption line
QSOs even much higher outflow velocities are observed (e.g. Turnshek 
et~al. 1988; Turnshek 1988)\nocite{tur88a,tur88}. However, there 
have not previously
been indications that such outflows originating from the AGN might continue 
at lower velocities out to distances of 70 kpc. Therefore, we regard it as 
unlikely that gas ejection by the AGN is the origin of the outer halo of 
1243+036 and its velocity shear.
\newline
An alternative outflow mechanism 
may be an equivalent of the `superwinds' observed in
low redshift strong IRAS galaxies (e.g. Heckman et~al. 1990; Heckman 
et~al. 1993\nocite{hec90,hec93b}).
These outflows are believed to be caused by a massive starburst in the galaxy.
In the burst of star formation, the kinetic energy supplied by 
stellar winds and supernovae results in the expulsion of interstellar material
into the surrounding medium, and can drive a large scale outflow.
Some ultraluminous IRAS galaxies (e.g. Arp\,220, Mrk\,266) have superwinds 
producing emission line nebulae several tens of kiloparsecs in size (see 
Heckman et~al. 1993\nocite{hec93b} and references therein). 
Linewidths and 
observed outflow velocities of emission line gas in 
the superwinds are typically a few hundred kilometers
per second, similar to what we find for the velocity shear and width of the 
Ly$\alpha$ emission in the outer halo of 1243+036. 
If a superwind is responsible for the presence and kinematics of the outer halo
Ly$\alpha$ gas in 1243+036, then the timescale of the outflow would mean that
a massive initial starburst must have occurred long before the radio source
switched on. Heckman et~al. 
(1990)\nocite{hec90} showed that
an outflow velocity $v_{100}= 100$ km s$^{-1}$ at a radius $r$ kiloparsec
with a density of the ambient medium $n_{ext}$ requires
a constant energy injection rate
$dE/dt \approx 3 \times 10^{41} r^2 v_{100}^3 n_{ext}$.
To produce this energy injection rate by a starburst requires a star formation
rate SFR$\sim (dE/dt)/(2.7 \times 10^{41}$ ergs s$^{-1}$) M$_{\odot}$ yr$^{-1}$.
The velocity shear and extent of the outer halo in 1243+036
implies an energy injection rate of
$\sim 10^{45}$ ergs s$^{-1}$ during a period of the halo's
dynamical timescale of $\sim 3 \times 10^8$ years. 
If this energy injection rate is indeed
produced by a starburst, it implies a star formation rate for 1243+036 of
$\sim 3000$ M$_{\odot}$ yr$^{-1}$ assuming a Salpeter IMF (Heckman et~al. 
1990 and references therein). A starburst that is 
biased towards more massive stars may still require 1000 M$_{\odot}$ yr$^{-1}$,
an order of magnitude more than derived for Arp220 (\cite{hec90}).
Over the dynamical timescale of the outer halo, 
the entire stellar population of a massive galaxy would have formed.
Although such star formations rates are enormous, they 
are also inferred from stellar population modelling of the colours of high 
redshift radio galaxies and are invoked
to explain the alignment
effect of the optical and radio axis of high redshift radio galaxies by 
jet-induced star formation (e.g. 4C\,41.17, Chambers et~al. 
1990\nocite{cha90}). Such a large population of young hot stars
could account for the photoionization of a large
part of the Ly$\alpha$ gas.

{\sl c) Infall}
\newline
The Ly$\alpha$ gas of the outer halo might be infalling gas from the
outer parts of the protogalactic halo. If the radio source
associated with 1243+036 is oriented at some angle to the plane of the sky,
gas infalling along the radio axis and anisotropically ionized by the AGN
could show a velocity structure
similar to that observed for the outer halo Ly$\alpha$ gas. 
However, the radio polarization data of 1243+036 show that
the southern radio structure is the least depolarized. If 1243+036 is embedded
in a magnetoionic halo, as appears to be the case for most radio galaxies
and quasars (Laing--Garrington effect, Garrington et~al. 1988; 
Laing 1988a; Carilli et~al. 1994\nocite{gar88,lai88,car94}), 
the southern radio jet should be on the side closest to us. The radiation from
the southern part of the radio source has then traveled a shorter distance 
through the magnetoionic halo than the the northern radio emission 
and is therefore the least depolarized.
This orientation is further supported by the blueshifted Ly$\alpha$ component
at the position of the southern radio knot B1, 
probably caused by the interaction of the 
radio jet and the gas (see Sect.~4.3). 
Therefore, it seems somewhat
surprising that ionized gas falling in from large radii on 
the southern side is also blueshifted with 
respect to the systemic Ly$\alpha$ and not redshifted.
Thus, the most likely orientation of
the radio axis with respect to the plane of the sky argues against the outer
halo Ly$\alpha$ emission being due to infall along the radio axis.
Nevertheless, we will consider whether an inflow mechanism could produce
the outer halo in 1243+036. 
\newline
There is much evidence that powerful radio galaxies at high redshifts reside
in clusters (e.g. Hill \& Lilly 1991\nocite{hil91} and references therein),
and that they have a hot ($\sim$10$^7$ K) ``cooling flow'' intra cluster 
medium (e.g. Cygnus A, Arnaud et~al. 1984\nocite{arn84}; 3C295, 
Henry \& Henriksen 1986\nocite{hen86}; 3C356, Crawford \& Fabian 
1993\nocite{cra93}; see also Fabian 1994\nocite{fab94}; Carilli et~al.
1994\nocite{car94}).
Our determination of the  minimum
pressures and polarization measurements of the radio structure of 1243+036
support this idea (see Sect.~4.1).
If the cooling time
of that gas is short compared to the age of the Universe at the observed epoch
then there will be a cooling flow, depositing cold gas near the centre of the halo.
At low redshifts the cooling time of the hot ICM in many clusters is shorter
than the Hubble time and therefore gas is cooling and deposited, possibly
forming emission line clouds. In nearby cooling flows mass deposition rates 
up to several hundred 
M$_{\odot}$ yr$^{-1}$ have been deduced (see Fabian 1994 and
references therein\nocite{fab94}).
\newline
The mass of the {\it ionized gas} in the outer halo of 1243+036 
in the observed area that was
covered by the slit in orientation PA1, 
is of order 10$^8$ M$_{\odot}$ (see Sect.~4.2). 
In the case of a cooling flow, the gas has no preferential direction of infall,
so that it is not only along the radio axis (slit orientation PA1), 
in the ionization cone from the 
AGN, but roughly isotropically distributed around the radio galaxy.
Thus, the total mass of cold ($\sim 10^4$ K)
gas in the outer halo all around the galaxy
is minimally of order 10$^9$ M$_{\odot}$, situated at a
radius of $\sim$60 kpc. If the ionized gas represents only a fraction of the
total mass, the total amount of gas in the outer halo
may be even an order of magnitude higher. At the redshift of
1243+036 the cooling time would have to be shorter than 10$^9$ years (the
approximate age of the Universe at that epoch) to allow a cooling flow to be
the origin of the emission line gas. From an estimated density of 0.1
cm$^{-3}$ at a radius $\sim$10 kpc (see Sect.~4.1), temperature $T\sim10^7$ K
and the cooling function
from Thomas (1988)\nocite{tho88}, the cooling time is $\sim$10$^8$ years. The
density in a hot halo decreases with radius at a rate between
$r^{-1}$ and $r^{-2}$, and the cooling time increases as $n^{-1}$ with
constant temperature (\cite{tho88}).
Then at $r=60$ kpc the cooling time is longer than in the inner part of
the Ly$\alpha$ halo, but is still $\sim$10$^9$ years. 
Thus it may be possible that a massive cooling flow has deposited
a large amount of gas in the outer halo of 1243+036.
\newline
A problem for the cooling flow hypothesis is the large velocity shear
of the outer halo. In a cooling flow, this is expected to be only a few tens of
kilometers per second (Fabian 1994; Fabian et~al. 
1987\nocite{fab94,fab87}), an order of magnitude smaller than that 
observed in 1243+036. A flow velocity of only 20 km s$^{-1}$ means a mass 
deposition
rate of $\sim$1000 M$_{\odot}$ yr$^{-1}$ within a radius of 60 kpc, 
assuming a density of 0.1 cm$^{-3}$.
Fabian et~al. (1987) suggests that galaxy 
interactions are most likely to cause large velocities and 
Keplerian motion of the gas deposited by a cooling flow.
Thus even though a cooling flow may be at least in part responsible for the
presence of the emission line gas in the outer halo of 1243+036, it cannot 
explain the observed velocity shear. An additional mechanism, e.g. a galaxy
interaction is needed to account for that.
\newline
However, unless our arguments about the orientation of the radio source are
wrong, i.e. the enhanced, blueshifted Ly$\alpha$ at the projected
position of bending point B1 is coincidental and the polarization asymmetry is
due to some extraordinary geometry of the magnetoionic halo, we regard the
infall of gas as the most unlikely explanation of the velocity shear.

In summary, we think that rotation due to accretion of gas associated with
the formation of the galaxy is the most plausible
origin for the presence and kinematics of the outer halo, but outflow in the
form of a `superwind' may also be feasible. A cooling flow may contribute to
the presence of the gas but cannot be the cause of the velocity shear. Infall
in general seems excluded by the most likely orientation of the radio source.
Regardless of the actual origin of the gas in the outer halo and its velocity
shear, it is clear that it must have been there since before 
(the currently active period of) the radio
source switched on.
\newline
Baum et~al. (1992) \nocite{bau92} classified low redshift radio galaxies
on the basis of the kinematics of the emission line gas into ``rotators'',
which Baum et~al. (1992) \nocite{bau92} associate with mergers,
``calm non-rotators'', identified with cooling flows, and ``violent 
non-rotators'', where
any organized motion of the emission line gas 
is overshadowed by interaction with the AGN.
1243+036 shows the
presence of two of the kinematical characteristics of the low redshift radio
galaxies: a calm rotator in the region outside the radio source and a violent
non-rotator in the inner parts. Thus, we 
may in 1243+036 be witnessing an important stage in the
evolution of massive galaxies in the early Universe. The scenario would be that
a large scale rotating gas disk or outflowing gas is ionized by the
strong continuum emission from the AGN, 
while the interaction of the radio source with the gas causes large
velocities removing signs of the earlier calm kinematics 
as the radio lobes propagate outwards. 
In the outer halo, where the gas is still unaffected by the
radio plasma, the rotating gas disk is still intact.
\newline
We note that 
in other high redshift radio galaxies a strong extended HI absorption feature
against the Ly$\alpha$ is frequently observed (\cite{oji95b,rot95a}). These
absorption features have velocity widths of a few hundred kilometers per second
and absorb a part of the emission line profile over the entire Ly$\alpha$
spatial extent along the radio axis. They may be caused by 
high column density
($\sim 10^{19}$ cm $^{-2}$) HI absorption systems with a Doppler parameter 
$b\sim 50$ km s$^{-1}$) or
by a superposition of many lower column density clouds with an integrated 
velocity dispersion like that observed of these absorption features.
If the latter explanation is the case,
the similar velocity width of the strong absorption features
observed in some HZRGs and of the Ly$\alpha$ emission from the
outer halo in 1243+036, suggests that these features may have a similar
origin.
\newline
Finally we note that the low surface brightness of the Ly$\alpha$ emission
from the outer halo of 1243+036 is of a similar level to that occasionally
observed from the extended nebulosity around radio quiet quasars (e.g. 
Bremer et al. 1992b)\nocite{brem92b} 
and from some damped Ly$\alpha$ absorption systems
(e.g. Pettini et al. 1995).\nocite{pet95} Perhaps this represents the quiescent
state of the gas around primeval galaxies in general, unaffected by
interaction with radio plasma.

\subsection{The bent radio structure and its relation with the Ly$\alpha$ gas}
Several authors have discussed 
multiple hotspots and bending of the radio jet in 
quasars and radio galaxies (e.g. van Breugel et~al. 1985;
Lonsdale \& Barthel 1986a, 1986b; Cox et~al. 1991; Icke 
1991).\nocite{bre85d,lon86,lon86a,cox91,ick91}
An important feature of the bending of the radio jet in 1243+036 is the 
coincidence of
enhanced Ly$\alpha$ emission that is at a blueshifted velocity with 
respect to the rest of Ly$\alpha$ emitting gas, suggesting a direct interaction
between the radio jet and the emission line gas. Line emission at the 
position of a sudden bend in the radio jet is also seen at low redshift in
3C\,277.3 (Coma A) (\cite{bre85d}).
We will consider three possibilities for the bent radio structure in 1243+036.

\noindent (i) {\sl Jet deflection by a gas cloud.}
\newline
Lonsdale \& Barthel (1986)\nocite{lon86} argue that the double hotspots 
often observed in powerful radio sources are due to deflection of the radio
jets by a gas cloud. They find that a straight deflection by a dense wall of a
massive cloud would be possible, but not very appealing, because very large
cloud masses are required ($>$10$^9$ M$_{\odot}$). Also, the jet would
quickly eat into the wall so that the deflection direction would change more
rapidly than the necessary time to build up the secondary hot spots.
\newline
Lonsdale \& Barthel (1986)\nocite{lon86} find that the best possible method
of deflection might be through a ``DeLaval'' nozzle (e.g. Blandford \& Rees
1974)\nocite{bla74}. The jet eats into a cloud, where it inflates a
plasma--bubble. The bubble breaks out at 
the weakest spot in the asymmetrically confining cloud,
creating the ``DeLaval'' nozzle. This then produces collimated
outflow that feeds the secondary hotspot. They claim that the cloud
with bubble structure will have a long lifetime such that the deflection
direction is maintained stationary over the time required to build up the
secondary hotspots ($\sim$10$^6$ years). The mass and density of the
deflecting cloud must be sufficiently large such 
that it is not blown away past the secondary hot spot.
\newline
From the total luminosity of 1243+036, the average energy supply rate to the
hot spots is 
unlikely to be larger than $\sim 3 \times 10^{45}$ ergs s$^{-1}$
(see Table 3).
From this and the total minimum energy of component B4,
we estimate that the time to build up component B4 is 10$^5$--10$^6$ years
(see also Lonsdale and Barthel, 1986a).
Thus, the advance speed ($v_h$) of hotspot B1 must be $<0.01c$ so that B1 does
not pass the other hotspots within that time.
From minimum energy density ($u_h$) for B1 in the
ram-pressure balance equation $v_h = u_h^{1/2} (3 n_{cl} m_{proton})^{-1/2}$,
the density of the deflecting cloud must be $>0.1$ cm$^{-3}$ (see also
Table 4), similar to what
Lonsdale and Barthel (1986) derived. 
This $0.1$ cm$^{-3}$ density gas represents the hot halo as suggested earlier
and is sufficient to provide the necessary time to build up the secondary
hotspots. As this hot gas most likely uniformly 
surrounds the whole system it does not itself
deflect the radio jet. It must be an overdensity of hot and cold gas, 
e.g. emission line gas, that deflects the radio jet. This whole region would
then be accelerated by the interaction and we shall refer to it as the 
deflecting ``cloud''.
\newline
The deflecting cloud must be larger than 
the size of the total working area of the radio jet, i.e. the primary hotspot.
The hotspot B1 is unresolved in our 8 GHz maps, i.e. smaller than 1.5 kpc.
The size of primary hotspots in powerful radio sources 
is typically 1 kpc (e.g. Laing 1988b).\nocite{lai88a} 
Assuming that the deflecting cloud has a diameter of $\sim$4 kpc, similar
to the determined for the enhanced Ly$\alpha$ emission region at B1, and
that it is spherical, the minimal density gives a total mass of
$\sim10^8$ M$_{\odot}$. The Ly$\alpha$ flux of the region of enhanced emission
at the position of B1 indicates the presence of $\sim10^7$ M$_{\odot}$ of
ionized gas (see Table 5). If the emission line clouds 
have dense neutral cores then the total mass of gas (neutral and
ionized) may be an order of magnitude higher. 
\newline
The Ly$\alpha$ image and high resolution spectroscopy show that the enhanced
emission at the location of radio hotspot B1 has a velocity of $\sim1100$ km
s$^{-1}$ with respect to the main Ly$\alpha$ gas. To determine whether the
interaction with the radio jet could have accelerated the gas we have to
estimate the jet power of 1243+036. 
A rough estimate can be made from the total radio
luminosity (see Table 3) and the conversion efficiency factor
$\epsilon$, 
L$_{jet}\sim10^{46}/2\epsilon$ (e.g. Bridle \& Perley 1984\nocite{bri84}), 
where the efficiency is probably of order 
0.1--0.01. Thus a jet power $>10^{46}$ ergs s$^{-1}$ is a reasonable assumption.
Alternatively we can estimate the jet power, assuming that the total minimum
energy of the radio components is replenished in the synchrotron lifetime,
giving L$_{jet}\sim 10^{47}$ ergs s$^{-1}$. A third estimate comes from
the jet thrust of a relativistic jet (not to be confused with the hot spot 
advance speed which is only $\sim 0.01 c$),
$\Pi_{jet}$=L$_{jet}(\gamma + 1)/(v_{jet}\gamma$) (\cite{bri84}), that can
be determined from the minimum pressure and area of a hotspot.
Assuming a hotspot size of 1 kpc (see above) and 
jet parameters like in Cygnus A, $\gamma=1.6$,
$v_{jet}\sim c$ (e.g. Perley et~al. 1984\nocite{per84a}),
we find L$_{jet}\sim 5$x10$^{46}$ ergs s$^{-1}$.
Thus it is reasonable to assume that the jet power of 1243+036 is $>10^{46}$
ergs s$^{-1}$.
A gas mass of 10$^8$ M$_{\odot}$ with a velocity of 1100 km s$^{-1}$ 
has a kinetic energy of 1.2x10$^{57}$ ergs. The jet can have accelerated this
mass over a period of 10$^5$ years, if the transfer efficiency of jet power 
to kinetic energy of the gas is $\sim$1\%.
\newline
In this deflection scenario
the pre-existing massive cloud that is hit by the radio jet does not
necessarily have to be one kiloparsec-scale cloud, which given the densities
of the emission line gas would be Jeans unstable,
but could consist of a region of
small dense clouds or filaments and low density gas in between. 
If the clouds are 
optically thick they can contain large amounts of neutral gas.
Whatever bends the radio jet has to be a continuous surface over the working 
area of the radio jet. Thus, if it is the emission line gas that deflects
the radio jet, the clouds must have been turned into a ``mist''.
The interaction with the radio jet may rip apart dense neutral cloud cores 
forming a
gaseous medium with an increased filling factor and covering fraction 
(\cite{brem95}).
This ``mist'' could be the medium in which the De Laval nozzle forms
by which the radio jet is delfected
The larger filling factor will cause a larger fraction of the gas to be exposed
to the ionizing radiation. This 
will enhance the Lyman $\alpha$ emission from the region. Local 
shock ionization due to the impact of the jet may further enhance the 
Ly$\alpha$ emission.
\newline
Alternatively Icke (1991) \nocite{ick91} showed that a jet can be bent
smoothly by a pressure gradient and thus may bounce off a gas cloud. 
In the case of 1243+036 the
redirected jet would then have several emission knots due to weak shocks from
further interactions with the emission line gas. 
Although this might account
for the observed radio structure of 1243+036 and the acceleration of emission
line gas at B1, it is not clear if this
mechanism would also cause the enhanced Ly$\alpha$ emission from the deflecting
cloud.

\noindent (ii) {\sl Precessing radio jet.} 
\newline
An alternative explanation for the bent radio structure with multiple hotspots
might be precession of the radio jet. The direction of the southern jet and
the slight northern extension of radio lobe A give the radio source an overall
Z--shape, suggesting that the jet may be rotating. As the tip of the jet moves
over the wall of the radio lobe it would 
create new hotspots, as in the ``dentist's
drill'' model of Scheuer (1982)\nocite{sch82}. The spectral steepening towards 
the outer parts
of the radio structure could be the result of synchrotron aging in older hot
spots. But it could also result from there being weaker shocks with distance
along the jet in the deflected flow model.
Our crude estimate of the synchrotron age of hotspot B4 of 10,000 years (see
Table 3) indicates that if the multiple hotspots are due to
the directional change of the radio jet, for B4 still to be visible when B1 is
being formed, the jet precession speed would be 0.001$^{\circ}$ yr$^{-1}$ and 
the jet would have revolved 27 times within the age of the radio source 
($\sim$10$^7$ yr).
The projected velocity of the jet impact site from B4 to B1 (10 kpc) 
would be $3c$.
At these precession speeds we would not expect the jet to be straight from
the core-jet component 
to B1 (and A1) and a straight line of hotspots from B1 to B4, but a more
gradually bent S--shape structure.
Also at this precession speed the time that the jet feeds the area of one 
hotspot is a factor 100--1000 shorter than the time required to build up 
the hotspot (see above). Nor would it have time to accelerate the emission
line gas at B1 to its observed velocity.
Even if we have underestimated the age of B4 by a factor of 10, the inferred 
speeds seem implausibly high.
\newline
Thus we regard the idea of a precessing jet causing the bent radio structure and
multiple hotspots as very unlikely.
Also the magnetic field projecting along the southern jet (Fig.~7)
is consistent with flux freezing and shear flow along the jet (\cite{beg84}),
and hence supports the idea of a deflected flow model.

We see the bent radio structure projected on the plane of the sky. However,
the bending
of the southern jet may have taken place partly into the direction
of the line of sight. If the jet is still moving relativistically after
the deflection, Doppler boosting may increase the observed flux of weak
radio knots along the jet. 
In the light of the orientation unification scheme for radio galaxies and
quasars, 1243+036 would be oriented close to the plane
of the sky. If the bending of the jet towards us is as much as it is
projected on the plane of the sky ($\sim30$ degrees), the southern jet could
have been deflected to as close as $\sim$30 degrees from our line of sight. For
a relativistically moving jet ($\gamma\sim$ a few) the observed flux could 
have been Doppler 
boosted by a factor $\sim$3--15 relative to its original direction. 
The synchrotron emission associated with 
weak shocks in the deflected southern jet may have been enhanced forming the
observed knots B2, B3 and B4.

\noindent (iii) {\sl Sweeping of the radio jet by the ambient gas.}
\newline
As a third possibility, the sudden bend and overall Z--shape of the radio source
may be due to  ``sweeping'' of the radio jet by the ambient medium 
after the jet has been slowed down in
the hotspots B1 and A1.  Such a mechanism has also been invoked
to explain the linear radio features in some Seyferts and the Z--shape of
the low redshift radio galaxy 3C\,293 (\cite{wil82,bre84a}) where the radio jets
are propagating through a rotating gaseous disk. 1243+036 shows both
morphological and kinematical similarities to 3C\,293 and 3C\,305 (\cite{hec82}) 
where there
is evidence for rotation and the radio morphology is Z--shaped.
At high redshifts not only 1243+036, but also 4C\,41.17 exhibits a radio 
morphology where the axis
defined by the outermost radio components is rotated with respect to the axis
defined by the inner components. These morphologies suggest that large scale
motion of the ambient medium may have deformed the outer parts of the radio 
source. The possible
rotating gaseous disk of the outer halo in 1243+036 indicates
that large scale organized motion is present that may provide the
transverse ram pressure (sweeping effect) that is required 
to bend the radio source to its observed morphology.
The velocity shear of the outer halo is consistent with solid-body rotation
that, if it is sweeping the radio plasma, would indeed cause the straight radio
structure of the ``southern jet'' in 1243+036.
Scaling the calculations of van Breugel et~al. (1984)\nocite{bre84a} for
3C\,293 to the much larger jet kinetic energy of 1243+036 (derived above),
requires an average
density $\sim2500\epsilon^{-1}$ cm$^{-3}$, where $\epsilon$ is the efficiency 
of order 0.1--0.01 as mentioned above, to bend the jet over a distance 
comparable to the jet diameter. Thus, the jet must
have lost more than 99.9\% of its kinetic energy in the hotspot B1 (and A1),
and continue onwards as a light, but still fast jet ($>10^4$ km s$^{-1}$, see 
van Breugel et~al. 1984\nocite{bre84a}),
to be bent by the ($\sim20$ cm$^{-3}$) rotating gas of the outer halo.
\newline
Although we cannot exclude this possibility, it seems more plausible that,
with a strong interaction at B1, the jet would be deflected along the lines
of scenario (i) than that it would continue at high velocity, but with 
much less thrust, in the same direction.

Thus, the bending of the radio structure in 1243+036 coinciding with accelerated
emission line gas is evidence that the radio jet is deflected 
by the interaction 
of the jet with the emission line gas. This supports the idea of
Barthel \& Miley (1988)\nocite{bar88b} that the
frequent distortions of radio structures of quasars and radio galaxies
at high redshifts are caused
by interaction of the radio jet with gas clouds in the environment.

\subsection{Aligned $R$ band continuum}
The $R$-band continuum morphology of 1243+036 is clearly aligned with the
principal radio axis, the line connecting components A, N and B1. This
alignment of $R$-band continuum emission with the radio axis is a commonly
observed feature in high redshift radio galaxies (\cite{cha87,mcc87a}). 
In addition to this, 1243+036 also has a
low surface brightness continuum component (see Fig.~8), 
which has the same narrow alignment
with the principal radio axis and apparently extends beyond the radio lobes.
If this faint extended continuum emission is
real, this can have important implications for the interpretation of the
alignment effect in high redshift radio galaxies.

One of the first proposed explanations for the ``alignment effect'' was jet
induced star formation (\cite{cha87,mcc87a,ree89a,beg89,you89}) where shocks
from the radio jet and the overpressured radio lobes would lead to star
formation in dense clouds along the direction of the jet propagation. 
Daly (1992) \nocite{dal92b} argues
that inverse Compton scattering of CMB photons of the relativistic electrons
in the radio plasma could cause the alignment effect at high redshifts.
Optical polarization measurements of intermediate and high redshift radio
galaxies have indicated that much of the aligned continuum light is polarized,
therefore some or all of the light is scattered light from a hidden quasar 
by dust or hot electrons (\cite{sca90,tad92,cim93,ser93,cim94,ser94}).
This indicates that, although jet
induced star formation may play a role, scattering processes
are important contributors to the alignment effect.
\newline
In 1243+036 we see that the aligned continuum does not follow
the radio structure that bends at the position of radio knot B1 into
the ``southern jet''. Instead the optical continuum extends
further outwards with the same orientation as the inner continuum, the
Ly$\alpha$ ``cone'' and the principal radio axis. This argues against a direct
relation between the optical continuum morphology and the radio jet as required
by inverse Compton models and jet-induced star formation and 
suggests that the optical continuum orientation is linked to the axis of the
AGN. However, in the jet-induced star formation scenario, 
the extended continuum
may be a relic of star formation of along a previous path of the jet, when
possibly no bending occurred.
Thus, the presence of faint extended optical continuum beyond the radio
structure maintaining its alignment with the inner radio structure and the
brightest Ly$\alpha$ gas favours the scenario of scattering of anisotropically
emitted nuclear continuum light, although it is possible that jet induced star
formation played a role at an earlier evolutionary stage of the radio source
and currently in the inner parts of the radio galaxy.
\newline
A much deeper image is needed to confirm the reality of the faint extended
continuum emission.

\section{Summary and Conclusions}
We have found and studied the distant radio galaxy 1243+036 at
$z=3.57$. It has spectacular properties both in the optical and in the radio:
\newline
1. The radio source extends over 55 kpc,
has a sudden bend and multiple hotspot structure. 
All components have ultra steep
spectra and the radio emission is strongly depolarized.
\newline
2. The radio galaxy has strong and extended Ly$\alpha$ and O[III] emission. The
Ly$\alpha$ emission is aligned with the radio source and shows a clumpy
structure. It is characterized by three components: (i) the inner halo
inside the radio structure with a large velocity dispersion, (ii) enhanced
emission with blueshifted velocity at the position of the bend in the radio
structure and (iii) a quiescent outer halo with a velocity shear of 450 km
s$^{-1}$ extending over a total diameter of at least 135 kpc, aligned with
the inner optical and radio axis.
\newline
3. The $R$ band continuum is aligned with the principal radio axis and shows a
faint extension slightly beyond the radio structure. This emission does not
have a sharp bend like the radio emission but stays aligned with the inner
continuum, the emission line gas and the principal radio axis.

We have argued that the observations have the following implications:
\newline
a) The quiescent outer Ly$\alpha$ halo must predate the onset of the radio
source. The origin of this gas and its kinematics is probably due to 
the accretion of gas from the environment
producing a large scale rotation, although there is a possibility that 
it is caused by 
a massive outflow from the central galaxy.
\newline
b) The interaction of the radio jet with the emission line gas is
responsible for the bent radio structure. The jet has accelerated the gas at the
interaction to a velocity of 1100 km s$^{-1}$ relative to the rest of the 
system and may have shred the emission
line clouds, locally enhancing the Ly$\alpha$ emission.
The bending of the radio structure in 1243+036 coinciding with accelerated
emission line gas is evidence that gas clouds may be the general cause
of bending in high redshifts radio sources.
\newline
c) The large velocity dispersion of the emission line gas inside the radio
structure and the low dispersion of the gas outside of it, support the idea
that the large velocity dispersions of the emission line gas in HZRGs is 
caused by entrainment of gas by the radio jet or shocks and 
turbulence in the radio plasma. 
\newline
d) The close alignment of the principal radio axis with the inner Ly$\alpha$
gas and especially with the Ly$\alpha$ from the outer halo argues for
photoionization by anisotropically emitted radiation from the AGN.
\newline
e) The extended Ly$\alpha$ emission suggests densities of the emission line 
filaments of order a few to 100 cm$^{-3}$.
The required confinement of the emission line gas and the radio hotspots 
along with strong depolarization of the radio 
emission argue for the presence of a dense (0.1 cm$^{-3}$) hot ($10^7$ K)
halo surrounding the system.
\newline 
f) The extension of the aligned optical continuum emission beyond 
the radio structure and 
not following the bending of the radio source is an argument 
against inverse Compton scattering and jet-induced
starformation for the alignment-effect in this
object. It favours scattering of anisotropic nuclear continuum radiation.

The most likely scenario for explaining our observations is that
a large rotating gaseous disk, originating from accretion 
associated with the formation of the galaxy, is ionized by anisotropically
emitted continuum emission from the AGN. The interaction of the radio source
with the gas bends the radio jet and 
causes large velocities in the Ly$\alpha$ halo
as the radio plasma propagates outwards. In the outer halo the gas is still
unaffected by the radio plasma and the quiescently rotating gas disk is still
intact.

Because of its large luminosity and spatial extent, 1243+036 provides an 
important laboratory for investigating the intrinsic nature of high reshift
radio sources and the conditions in the early Universe. Further observations
of this object are underway using ground based telescopes and the Hubble 
Space Telescope.

\section*{Acknowledgements}

We acknowledge support from an EEC twinning project and a programme subsidy
granted by the Netherlands Organization for Scientific Research (NWO).
We thank Steve Eales and Steve Rawlings for their kindness to obtain a
$K$ band image.


{\small
\begin{table*}
{{\bf Table 1.} The Observations.}
\begin{center}
\begin{tabular}{ccccc}  \hline
Date        &  Telescope   & Frequency/Wavelength & Integration-time & Resolution \\[0.1cm] \hline
26 Nov 1988 &  VLA A-array & 1465, 1515  MHz         & 300 s   & $1''$   \\
20 Aug 1991 &  VLA A-array & 1465, 1515 MHz         & 2520 s   & $1''$   \\
20 Aug 1991 &  VLA A-array & 8415, 8465 MHz         & 7200 s   & $0.23''$   \\
18 Mar 1994 &  VLA A-array & 8085, 8335 MHz   & 1200 s   & 0.23$''$ \\
18 Mar 1994 &  VLA A-array & 4535, 4885 MHz     & 600 s   & 0.43$''$ \\
01 Mar 1990 &  ESO 2.2m    & $R$ band & 2700 s & $1.0''$ \\
21 Mar 1991 &  ESO NTT + EMMI & 4100 to 7900 \AA    & 7200 s(PA1) & $1.5''$ x 12 \AA \\
14 Apr 1994 &  ESO NTT + EMMI & 5300 to 5900 \AA    & 14400 s(PA1) & $1''$ x 2.8 \AA \\
14 Apr 1994 &  ESO NTT + EMMI & 5300 to 5900 \AA    & 3600 s(PA2) & $1''$ x 2.8 \AA \\
16 Apr 1994 &  ESO NTT + SUSI & 5571 \AA/60 \AA & 7200 s & $0.6''$ \\
14 Jun 1994 &  UKIRT + CGS4 & 2.2 to 2.4 micron & 3600 s & $1.5''$ x 0.006 micron \\ \hline
\end{tabular}
\end{center}
NOTE-- PA1 is slit position angle of 152$^{\circ}$ (along the radio axis), PA2 is 62$^{\circ}$
\end{table*}
}


{\small
\begin{table*}
{{\bf Table 2.} Results from the radio polarimetric imaging of 1243+036}
\begin{center}
\begin{tabular}{@{\extracolsep{-0.3em}}cccccrrccc}  \hline
Comp. &  I$_{8.3}$ & I$_{4.7}$  & S$_{8.3}$ & S$_{4.7}$ & R.A.~~~~ & Decl.~~~~ & $\alpha$  & FPol$_{8.3}$ & 
D$_{8.3}^{4.7}$ \\ 
    & {\small mJy/beam} & {\small mJy/beam} & mJy & mJy & J2000~~~~ & J2000~~~~ & ~ & $\%$ & ~ \\ \hline
A2  & 0.95     & ~        & 2.1 &     & 12$^h$ 45$^m$ 38.28$^s$ & 03$^{\circ}$ 23$'$ 23.1$''$ & ~ & ~ & ~ \\
A1  & 3.25     & 10.62    & 5.2 &20.3 & 38.29$^s$ & 22.6$''$ & $-1.7$ & 4.4 & $<$0.30$\pm0.1$ \\
N   & 0.31     & 0.70     & 0.7 & 1.4 & 38.37$^s$ & 21.0$''$ & $-1.0$ & ~ & ~ \\
B1  & 9.87     & 19.19    & 11.8&22.2 & 38.43$^s$ & 19.5$''$ & $-1.0$ & $<$2.0 & ~ \\
B2  & 1.49     & 5.91     & 2.8 &11.6 & 38.43$^s$ & 18.8$''$ & $-1.6$ & 17 & 0.55$\pm0.1$ \\
B3  & 0.93     & ~        & 1.8 &     & 38.42$^s$ & 18.4$''$ & ~ & ~ & ~ \\
B4  & 1.80     & 8.56     & 3.8 &14.1 & 38.42$^s$ & 17.4$''$ & $-2.0$ & 0.30 & 1.3$\pm0.5$  \\ \hline
\end{tabular}
\end{center}
NOTE1-- I is the peak surface brightness. S is the integrated flux measured 
in a rectangular box around the component. $\alpha$ is the 
spectral index between 4.7 GHz and 8.3 GHz determined from the surface 
brightness after concolving the
maps to the same resolution. FPol$_{8.3}$ is the fractional polarization
at 8.3 GHz and D$^{4.7}_{8.3}$ is the depolarization of the radio
emission between 4.7 and 8.3 GHz.
\newline
NOTE2-- The components A1, A2 and components B2, B3 are unresolved at 4.7 GHz.
Values at this frequency stated only for A1 and B2 are actually for A1+A2 and
B2+B3.
\end{table*}
}


\begin{table*}
{{\bf Table 3.} Physical Parameters from the radio emission of 1243+036}
\begin{center}
\begin{tabular}{@{\extracolsep{-0.2em}}ccccccccc}  \hline
Comp. & P$_{4.7\ GHz}$ & L$_{tot}$ & B$_{min}$ & E$_{min}$ & Minimum & $P_{min}$ & $\nu_{break}$ & t$_{synchr}$ \\ 
          & x $10^{27}$ & x $10^{45}$ & x $10^{-4}$ & x $10^{-9}$ 
& Total Energy & x $10^{-9}$ &    &   \\
      & W Hz$^{-1}$    & ergs s$^{-1}$ &  Gauss  & ergs cm$^{-3}$ 
& $10^{58}$ ergs & dyne cm$^{-2}$ & GHz & 10$^3$ yr \\ \hline
Integrated & 11  & 7.1 &    &     &      &     & & \\
A2 &      &      & 24 & 540 & 13.7 & 180 & & \\
A1 & 2.2  & 1.7  & 11 & 116 &  3.0 &  39 & $\sim24$ & $\sim5$\\
N  & 0.07 & 0.06 & 3 &   9 &  0.2 &   3 &  &  \\
B1 & 1.9 & 0.9 & 8 &  62 &  0.7 &  21 & $\sim120$ & $\sim5$ \\
B2 & 0.9 & 1.0 & 16 & 227 &  2.6 &  76 &  & \\
B3 &     &     & 14 & 173 &  2.0 &  58 & & \\
B4$^1$ & 1.5 & 1.2 & 11 & 112 & 1.3 & 37 & $\sim25$ & $\sim10$ \\ \hline
\end{tabular}
\end{center}
$^1$ Minimum energy requirements for B4 were derived using the overall spectral index $-1.3$,
because the high frequency spectral index $-2.0$ would largely 
overestimate its total luminosity.
\newline
NOTE1-- The components A1, A2 and components B2, B3 are unresolved at 4.7 GHz.
Values at this frequency stated only for A1 and B2 are actually for A1+A2 and
B2+B3.
\newline
NOTE2-- For calculation of the luminosity of the components with high frequency
spectrum steeper than $-1.0$, a two-slope power law spectrum was assumed with a
spectral index below 4.7 GHz (observed frame) of $-1.3$, equal to the integrated
source spectral index.
\end{table*}


\begin{table*}
{{\bf Table 4.} Density of the external medium from confinement by Ram 
pressure and static thermal pressure}
\begin{center}
\begin{tabular}{ccc}  \hline
Component & (n$_e$)$_{ext}$ & (n$_e$)$_{ext}$ \\ 
          & Ram $v$=3x10$^3$ km s$^{-1}$  & T=10$^7$K\\
          & cm$^{-3}$ &  cm$^{-3}$ \\ \hline
A2 & 1.2 & 56  \\
A1 & 0.3 & 12  \\
N  & 0.02& 1.0  \\
B1 & 0.1 & 6.6  \\
B2 & 0.5 & 24  \\ 
B3 & 0.4 & 18  \\
B4 & 0.3 & 12  \\ \hline
\end{tabular}
\end{center}
\end{table*}


\begin{table*}
{{\bf Table 5.} Density and mass of the ionized gas assuming $f_v=10^{-5}$}
\begin{center}
\begin{tabular}{ccccc}  \hline
Ly$\alpha$ Component & Size    & $L_{Ly\alpha}$           & (n$_e$)   & Mass \\ 
                     & kpc$^3$ & 10$^{44}$ ergs s$^{-1}$  & cm$^{-3}$ & 10$^8$ M$_{\odot}$ \\ \hline
inner halo           & 55x27x27&   2.7                    & 80        &  7   \\
central peak         & 14x6x6  &    1                     & 400       & 0.4  \\
peak at B1           & 4x4x4   &  0.3                     & 650       & 0.1  \\
outer halo           & (135-55)x27x27& 0.3                & 20        & 2.8  \\ \hline
\end{tabular}
\end{center}
\end{table*}

\hfill

\newpage \clearpage 

\begin{figure*}
\hspace{1.5cm}\hbox{
\psfig{figure=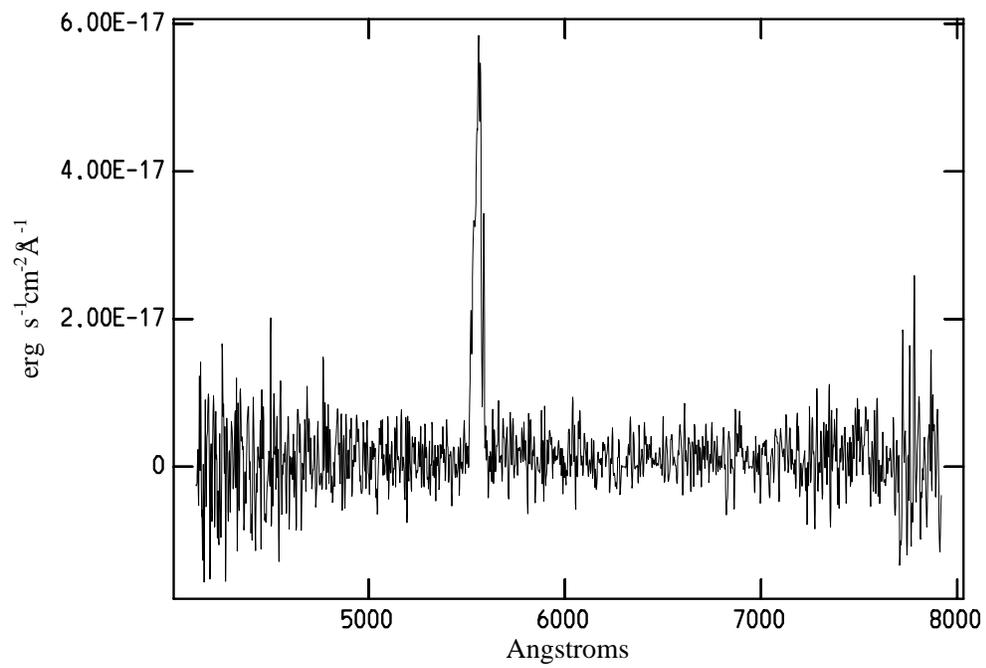,angle=90,width=14cm}}
\caption{ \label{lowresLya} 
The low resolution optical spectrum showing
one bright emission line, identified with Ly$\alpha$
}
\end{figure*} \newpage \clearpage 

\begin{figure*}
\hspace{1.5cm}\hbox{
\psfig{figure=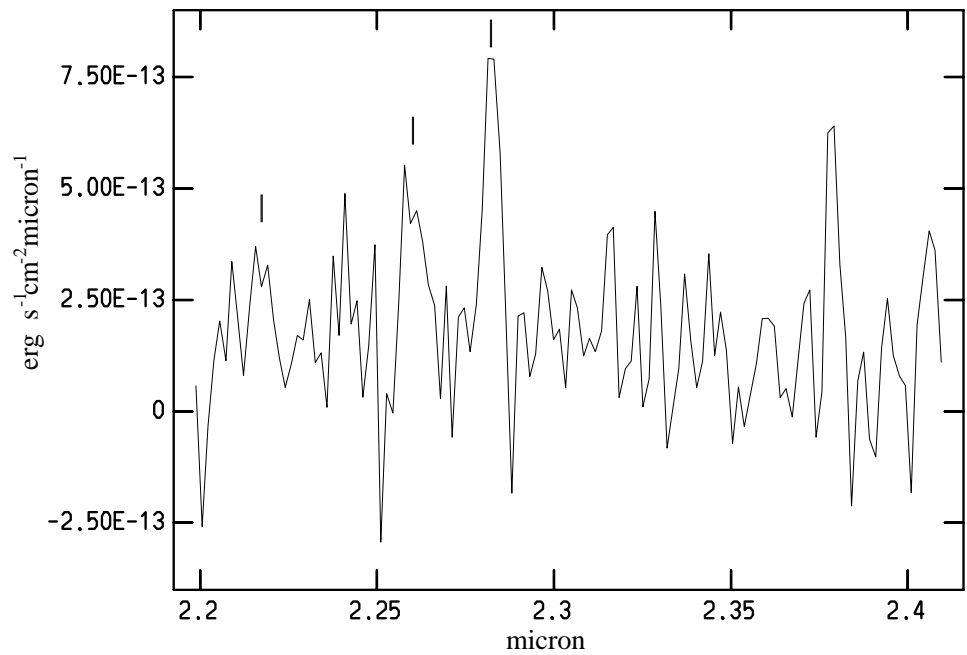,angle=90,width=14cm}}
\caption{ \label{IRspec} The infrared spectrum showing the redshifted 
[OIII] $\lambda$ 5007 \AA\ and 4959 \AA\ emission
lines and possibly H$\beta$ (at 2.22 micron). 
The feature at 2.38 micron is not real, but 
due to a residual hot pixel on the array.
}
\end{figure*} \newpage \clearpage 

\begin{figure*}
\hbox{
\psfig{figure=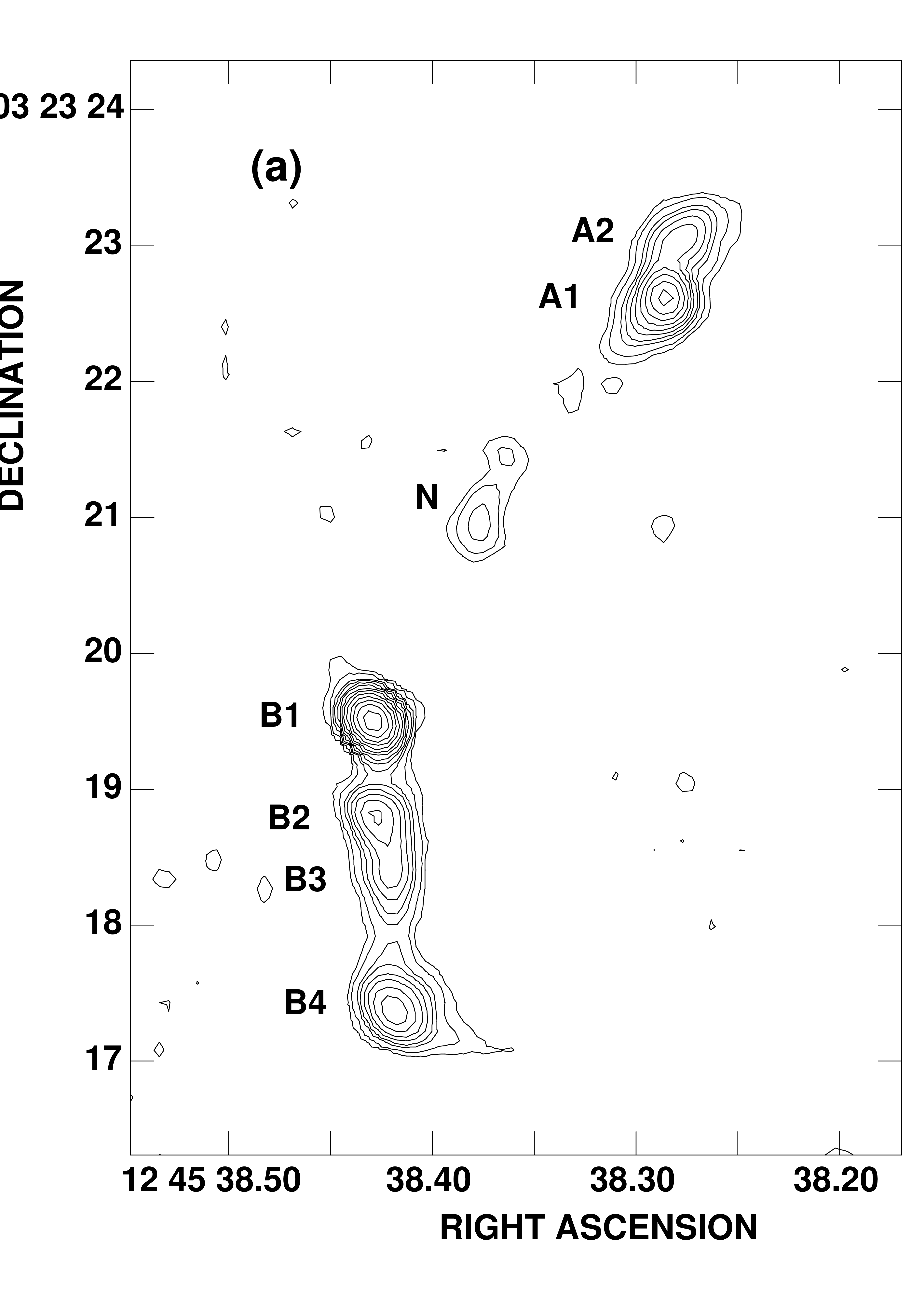,width=8cm}
\psfig{figure=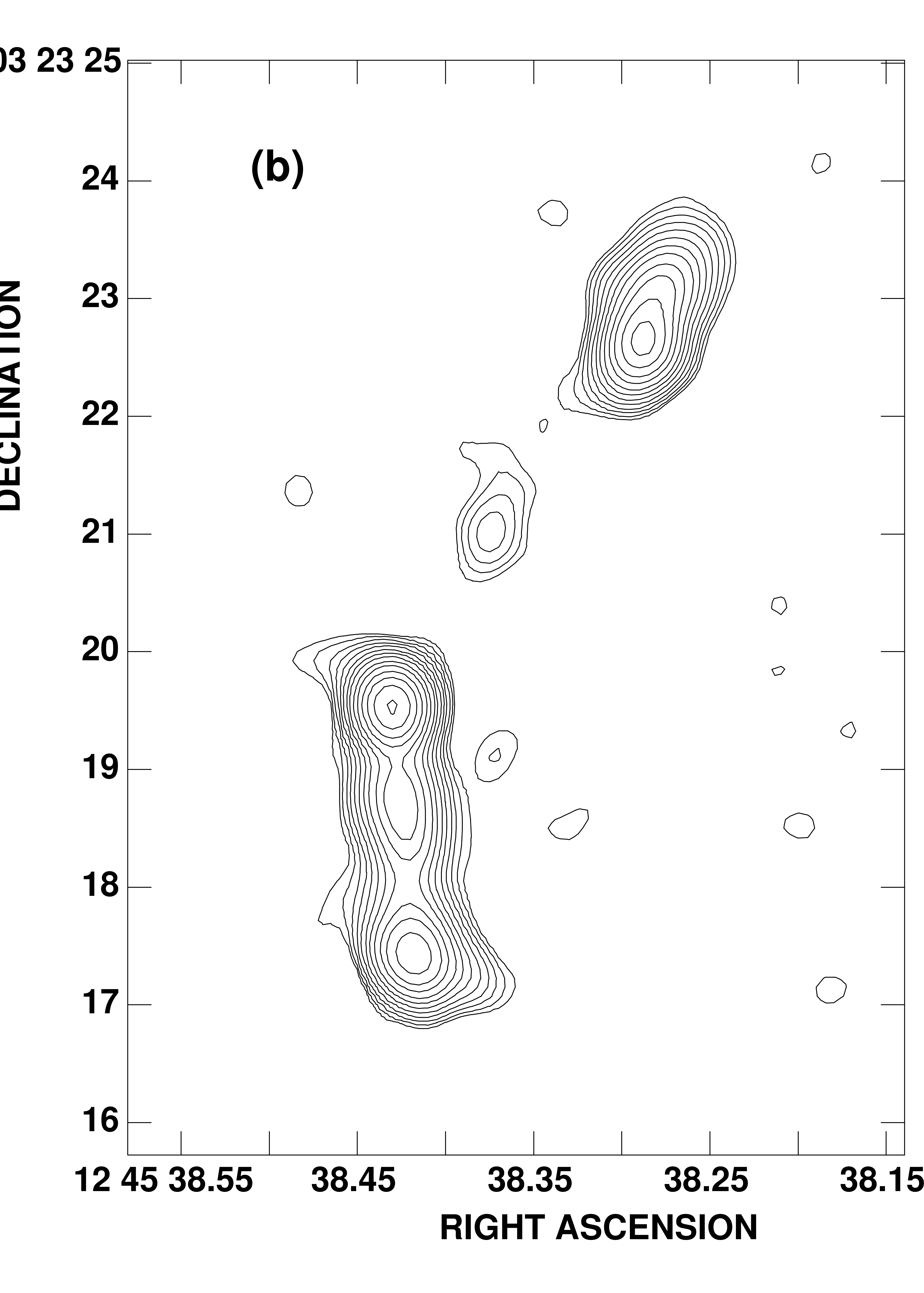,width=8cm}}
\caption{ \label{VLAim}  
 {\bf a}
An image of 1243+036 at 8.3 GHz with a Gaussian
restoring beam of FWHM = 0.23$''$. The contour levels are a geometric
progression in 2$^{1/2}$. The first contour level is 0.13 mJy, which is
three times the off-source RMS on the image. Three negative contours are
included (dotted).
{\bf b} An image of 1243+036 at 4.7 GHz with a Gaussian
restoring beam of FWHM = 0.43$''$. The contour levels are a geometric
progression in 2$^{1/2}$. The first contour level is 0.2 mJy, which is
three times the off-source RMS on the image. Three negative contours are
included (dotted).
}
\end{figure*} \newpage \clearpage 

\begin{figure*}
\centerline{
\psfig{figure=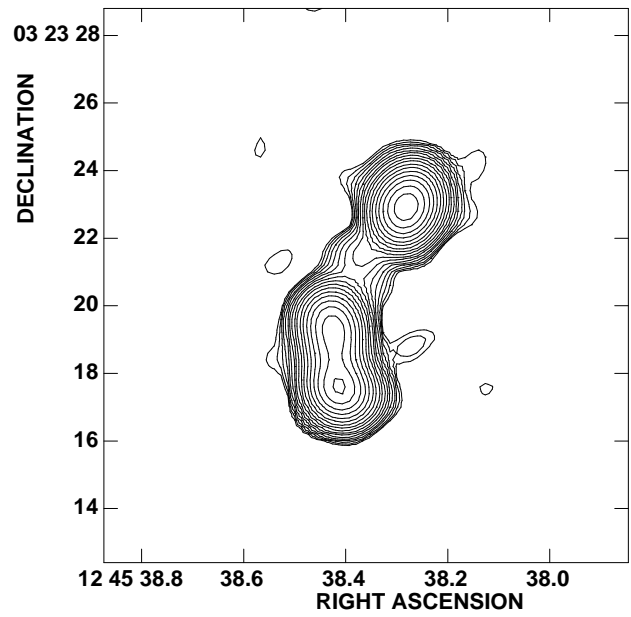,width=8cm}}
\caption{ \label{VLAim1.5} An image of 1243+036 at 1.46 GHz with a Gaussian
restoring beam of FWHM = 1$''$. The contour levels are a geometric
progression in 2$^{1/2}$. The first contour level is 0.3 mJy, which is
three times the off-source RMS on the image. Three negative contours are
included (dotted).
}
\end{figure*} \newpage \clearpage

\begin{figure*}
\hbox{
\psfig{figure=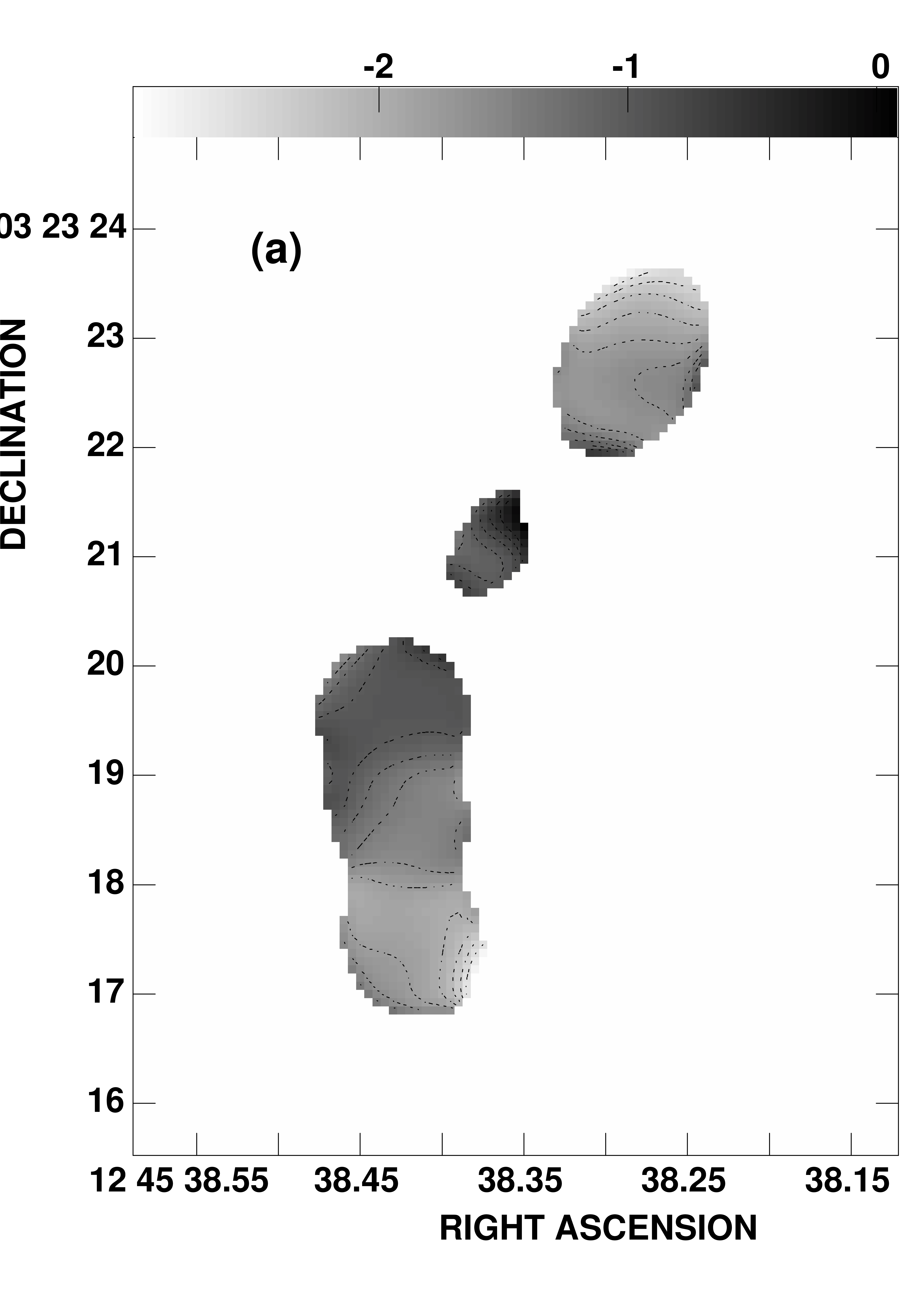,width=8cm}
\psfig{figure=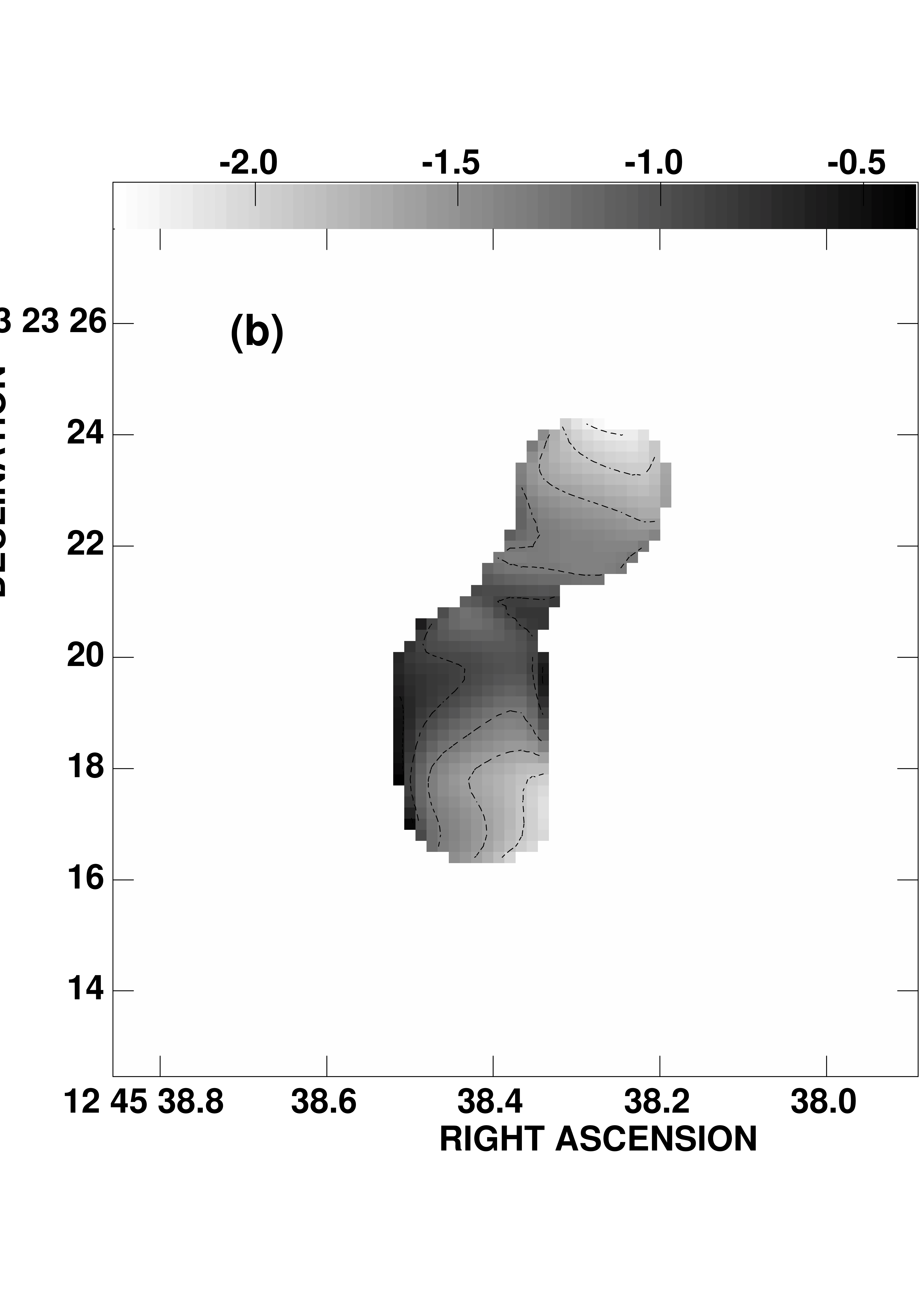,width=8cm}}
\caption{ \label{spix} 
 {\bf a} 
An image of the spectral index distribution
of the radio emission from 1243+036 at $0.43''$ resolution between
8.3 GHz and 4.7 GHz. The contour levels are: -2.8, -2.4, -2.2,
-2.0, -1.8, -1.6, -1.4, -1.2, -1.0, -0.8, -0.4 and -0.2. The grey-scale ranges
from -3 (white) to -0.3 (black).
{\bf b} An image of the spectral index distribution
of the radio emission from 1243+036 at $1''$ resolution between
1.46 GHz and 4.7 GHz. The contour levels are:  -2.2,
-1.9, -1.6, -1.3, -0.9, -0.6, -0.3 and 0.0. The grey-scale ranges
from -2.3 (white) to -0.3 (black)
}
\end{figure*} \newpage \clearpage

\begin{figure*}
\hbox{
\psfig{figure=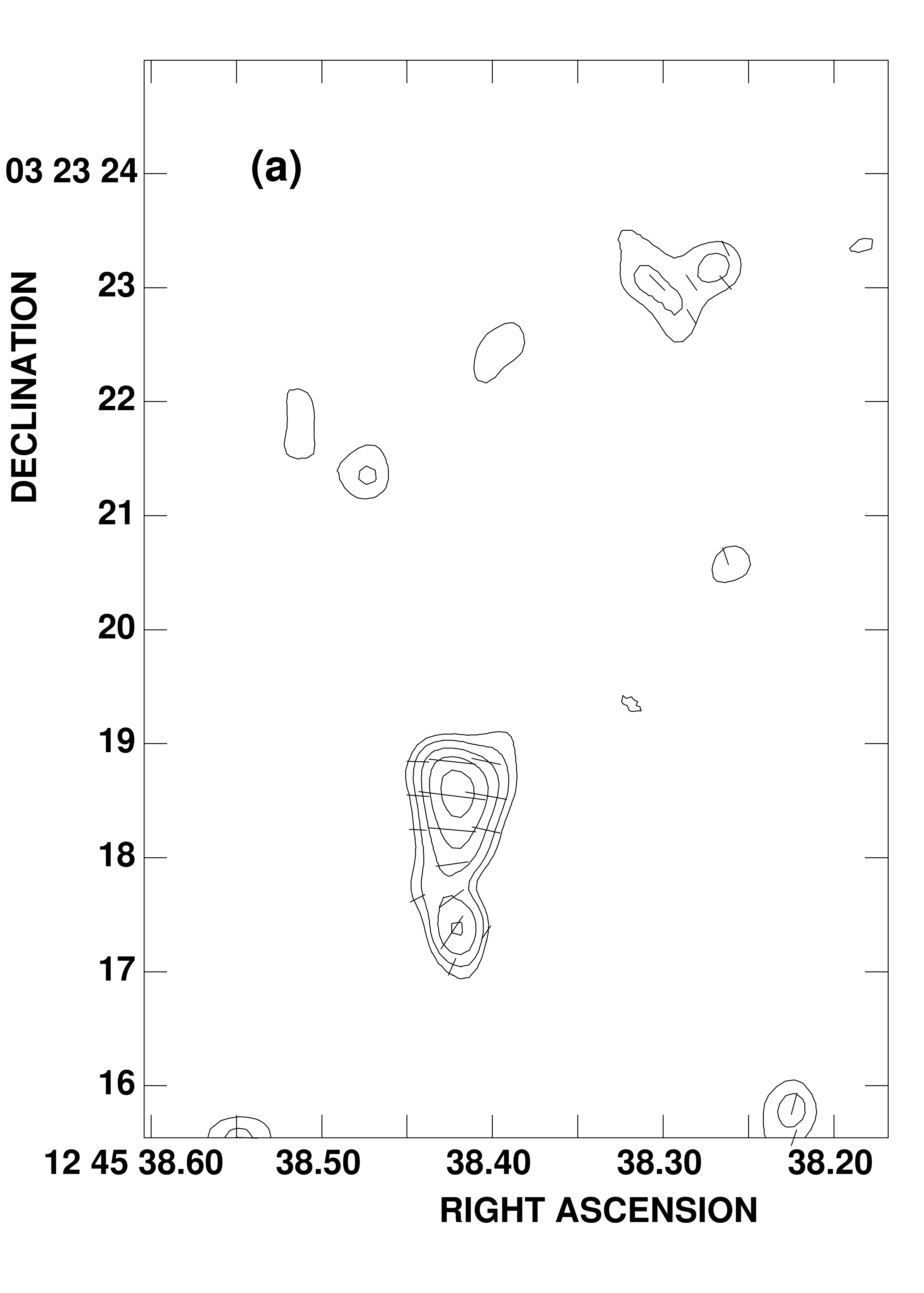,width=8cm}
\psfig{figure=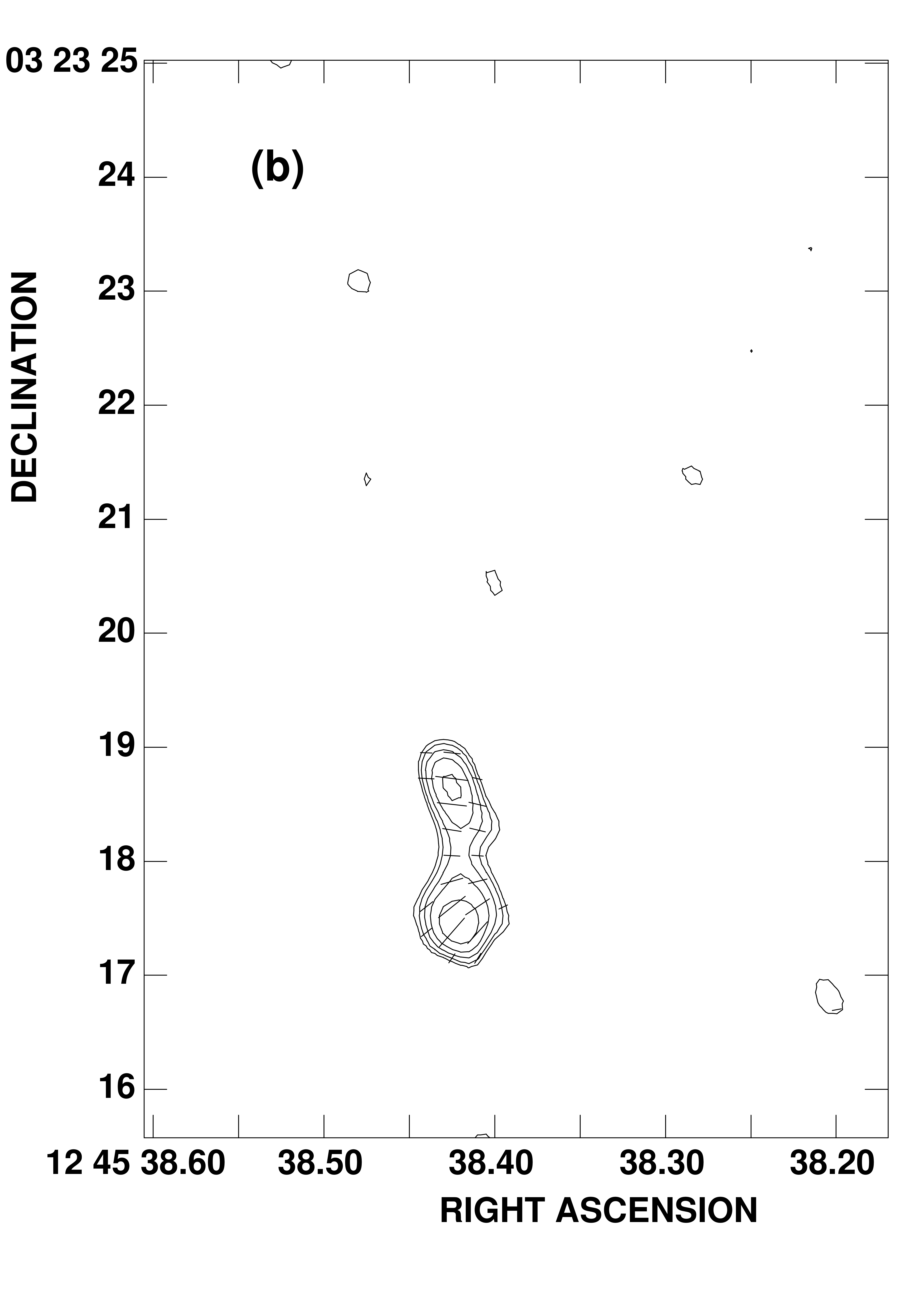,width=8cm}}
\caption{ \label{polarization} 
 {\bf a} An image of linear polarized intensity from
1243+036 at 8.3 GHz with a resolution of 0.43$''$. The contour levels
are a geometric progression in 2$^{1/2}$. The first contour level is 0.11 mJy.
The line segments show the orientation of the electric vectors
for the polarized emission.
{\bf b} An image of linear polarized intensity from
1243+036 at 4.7 GHz with a resolution of 0.43$''$. The contour levels
are a geometric progression in 2$^{1/2}$. The first contour level is 0.18 mJy.
The line segments show the orientation of the electric vectors
for the polarized emission.
}
\end{figure*} \newpage \clearpage

\begin{figure*}
\centerline{
\psfig{figure=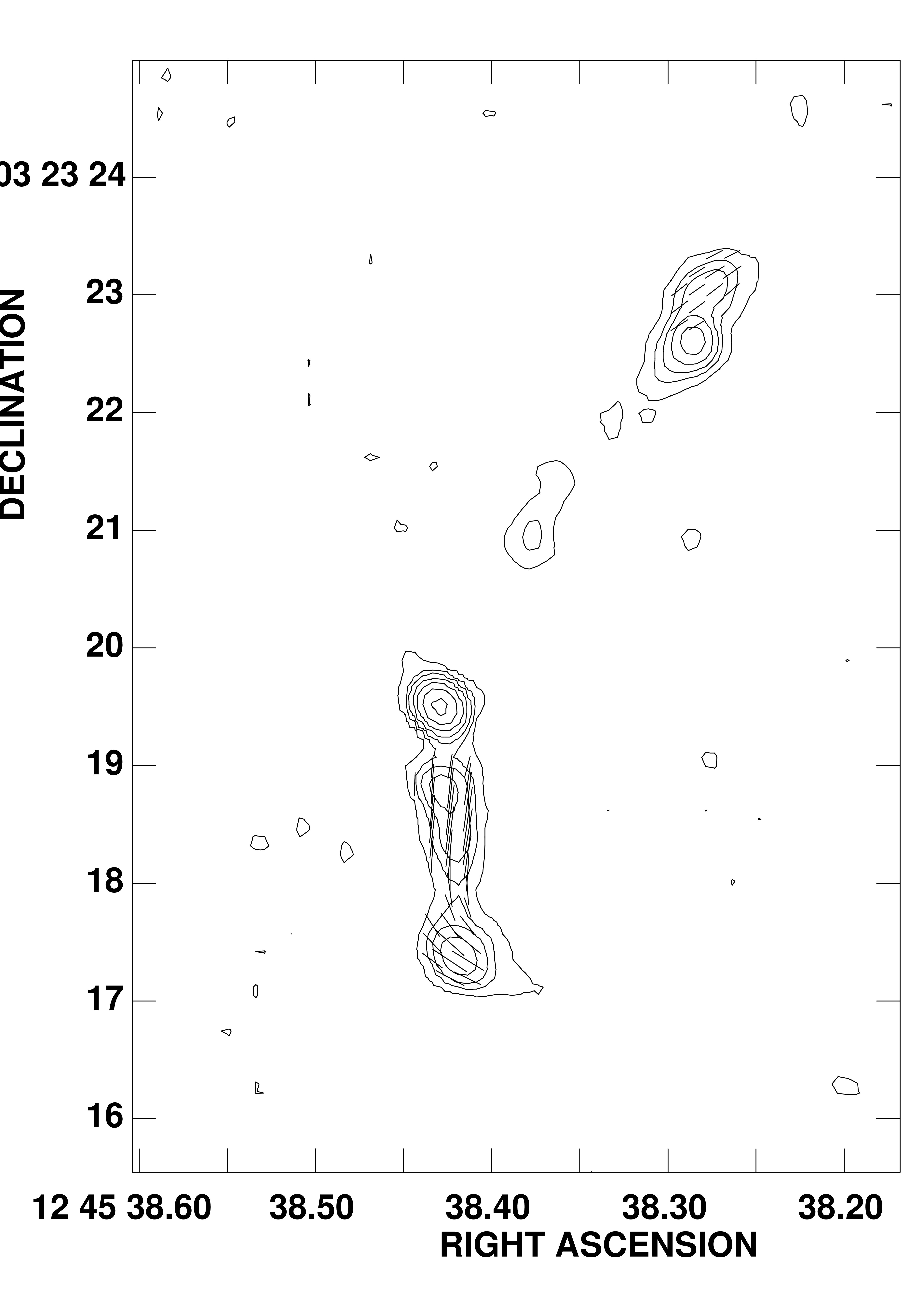,width=8cm}}
\caption{ \label{Bfield} The contours show the total intensity radio emission
from 1243+036 at 8.3 GHz. Contours are at 0.03 mJy x (-4, 2, 8, 16, 32, 64,
256, 512).
The line segments show the orientation of the projected magnetic field vectors,
derived from the polarized emission.
}
\end{figure*} \newpage \clearpage

\begin{figure*}
\centerline{
\psfig{figure=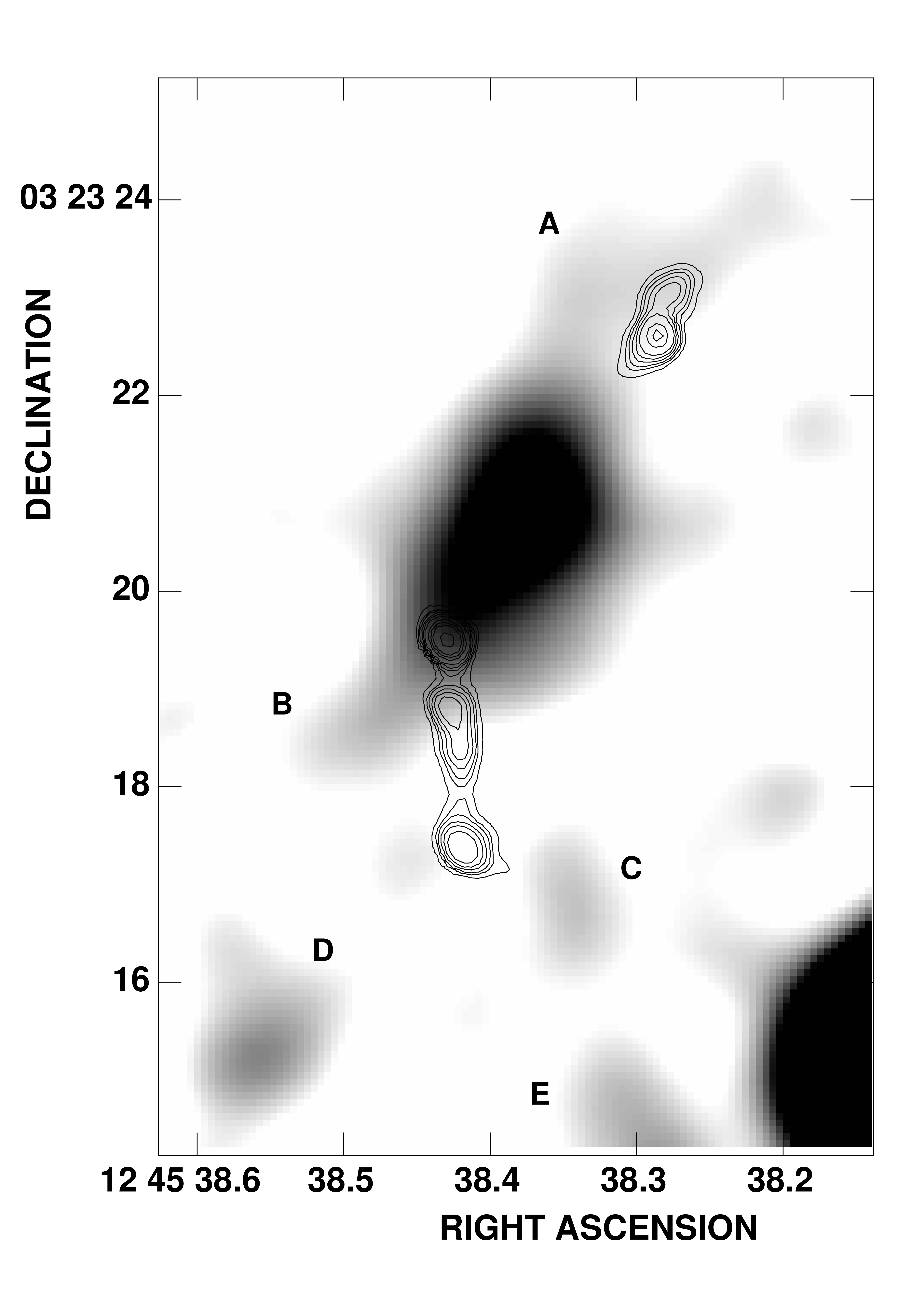,width=8cm}}
\caption{ \label{Rbandsm} 
Smoothed $R$ band image from R\"ottgering
$et$ $al.$ (1995b) with the contours of the 8.3 GHz radio image overlayed.
}
\end{figure*} \newpage \clearpage 

\begin{figure*}
\centerline{
\psfig{figure=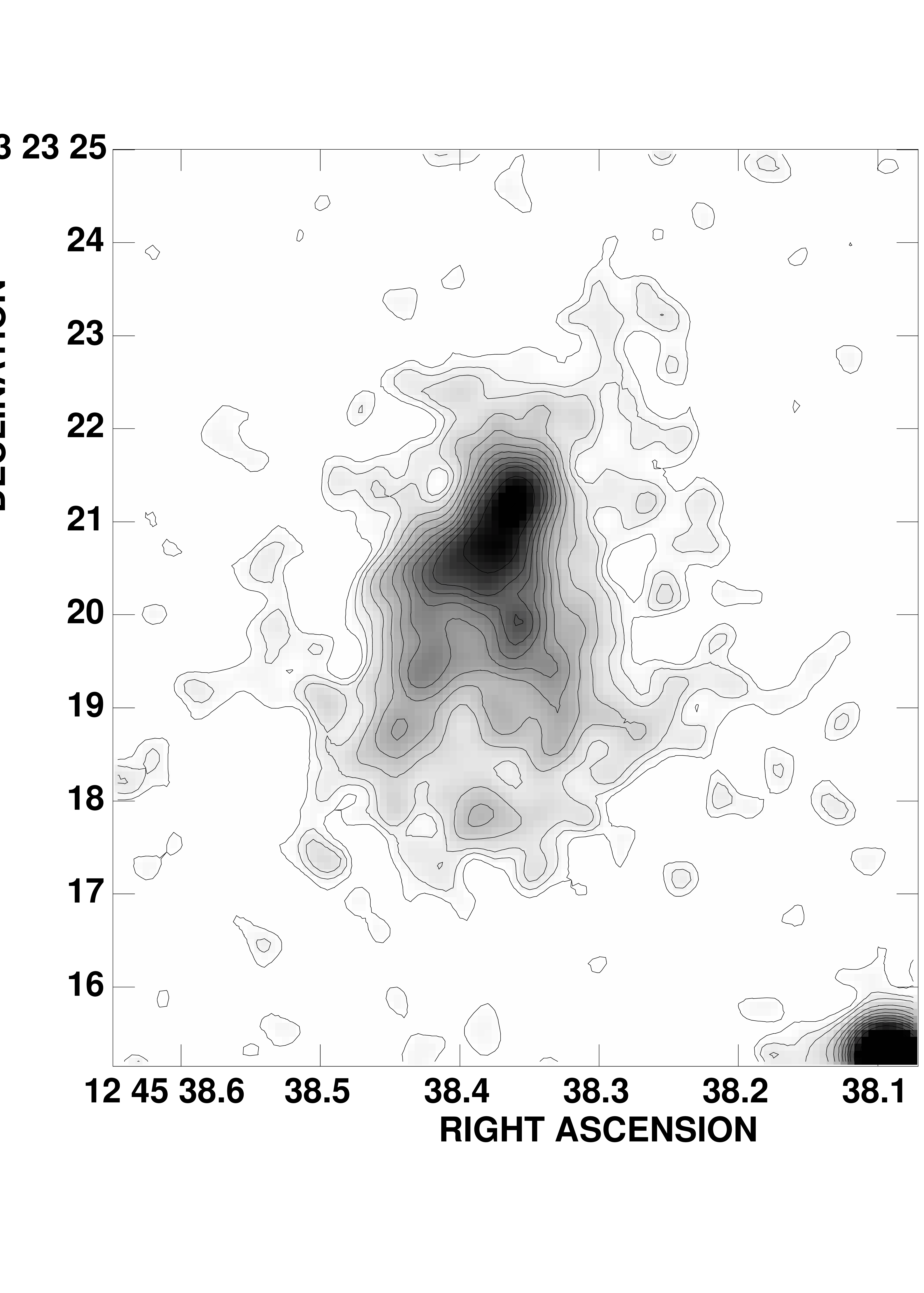,width=8cm}}
\caption{ \label{Lya} Image of the Ly$\alpha$ emission associated with
1243+036 with a resolution of $0.6''$. Contours linearly spaced are at 
2$\sigma$, 4$\sigma$, 6$\sigma$,..., where $\sigma$ is the background rms 
noise.
}
\end{figure*} \newpage \clearpage 

\subsection{High resolution spectroscopy}
\begin{figure*}
\hspace{1.5cm}\hbox{
\psfig{figure=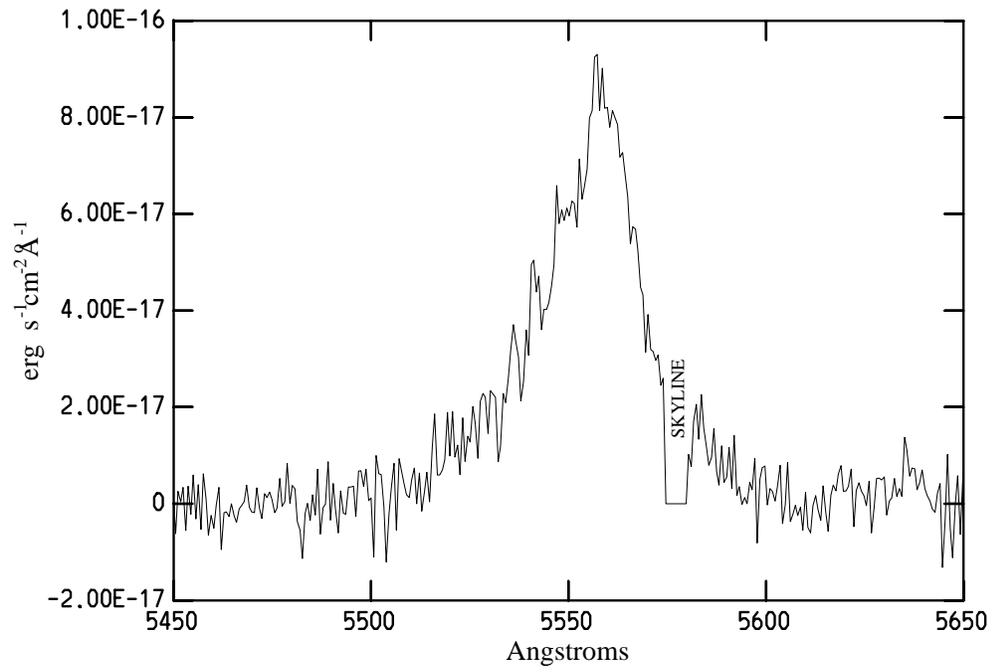,angle=90,width=14cm}}
\caption{ \label{Lyaprof1} The spectral profile of the central Ly$\alpha$ 
emission at high resolution (2.8 \AA) 
in position angle along the radio axis (PA1).
Details on the profile are described in the text.}
\end{figure*} \newpage \clearpage

\begin{figure*}
\hspace{-3cm}\hbox{
\psfig{figure=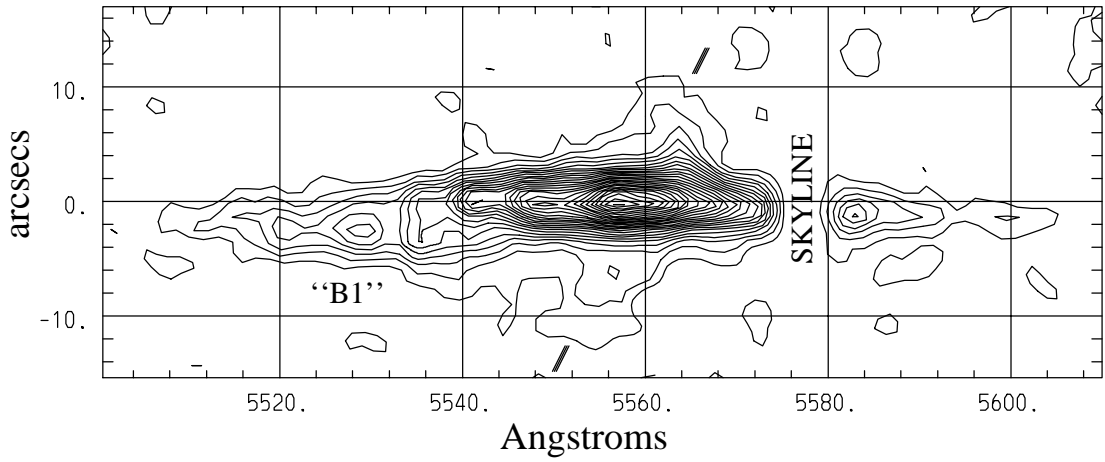,angle=90}}
\caption{ \label{2dimLya} A two-dimensional representation of the high
resolution spectrum of Ly$\alpha$ taken in PA1. 
The spectrum has been smoothed with a Gaussian of FWHM $1.1''$. 
The Ly$\alpha$ component
with a blueshifted velocity of 1100 km s$^{-1}$ is marked ``B1''. We
have further marked the narrow width Ly$\alpha$ emission with a velocity 
shear and an extent of more than 20''. Contours are linearly spaced 
at 2$\sigma$, 4$\sigma$, 6$\sigma$,..., where $\sigma$ is the background rms 
noise.
}
\end{figure*} \newpage \clearpage

\begin{figure*}
\hspace{-3cm}\hbox{
\psfig{figure=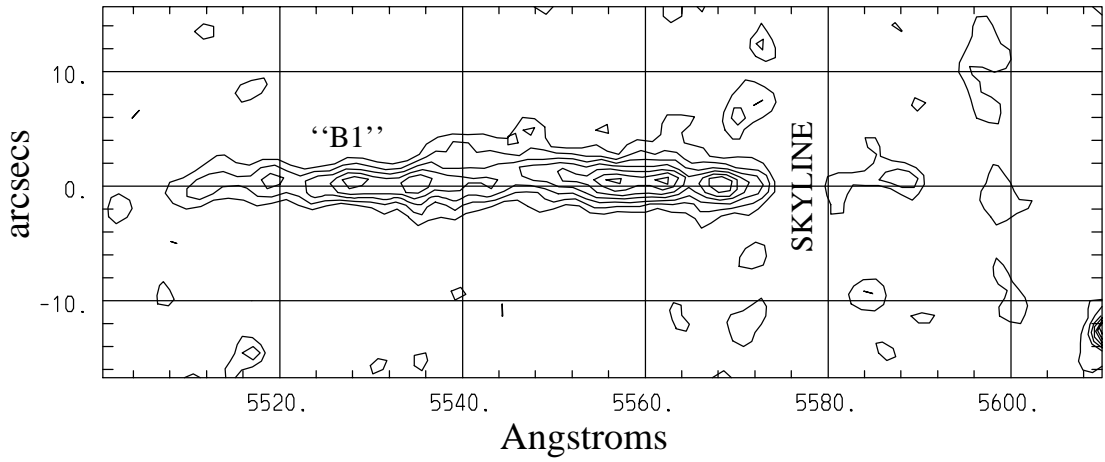,angle=90}}
\caption{ \label{2dimLyaPA2} A two-dimensional representation of the high
resolution spectrum of Ly$\alpha$ taken in PA2. The spectrum has been smoothed
with a Gaussian of FWHM $1.1''$.
The Ly$\alpha$ component
with a blueshifted velocity of 1100 km s$^{-1}$ is also seen in this 
spectrum and marked ``B1''.
Very extended Ly$\alpha$ emission with a velocity shear and narrow width
as seen in the spectrum along the radio axis is not seen along PA2.
Contours are linearly spaced at 2$\sigma$, 3.5$\sigma$, 5$\sigma$,..., where
$\sigma$ is the background rms noise.
}
\end{figure*} \newpage \clearpage 

\begin{figure*}
\hspace{1.5cm}\hbox{
\psfig{figure=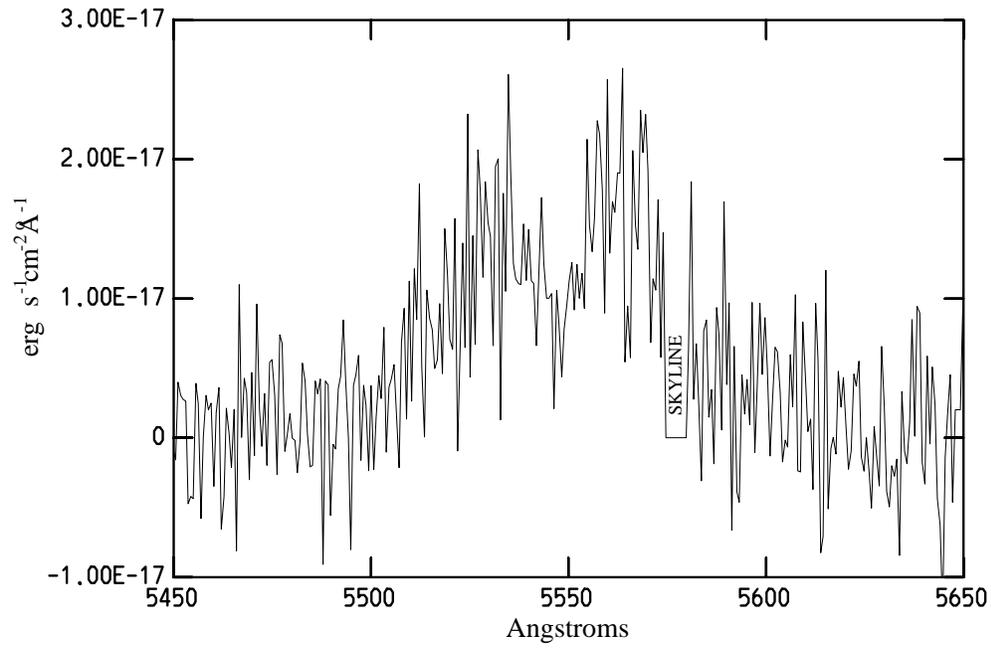,angle=90,width=14cm}}
\caption{ \label{Lyaprof2} The spectral profile of the central Ly$\alpha$
emission at high resolution (2.8 \AA)
in position angle perpendicular to the radio axis (PA2).
The slit was positioned slightly off the brightest central Ly$\alpha$ emission, 
therefore including emission from component B1 (5530 \AA).}
\end{figure*} \newpage \clearpage 

\begin{figure*}
\centerline{
\psfig{figure=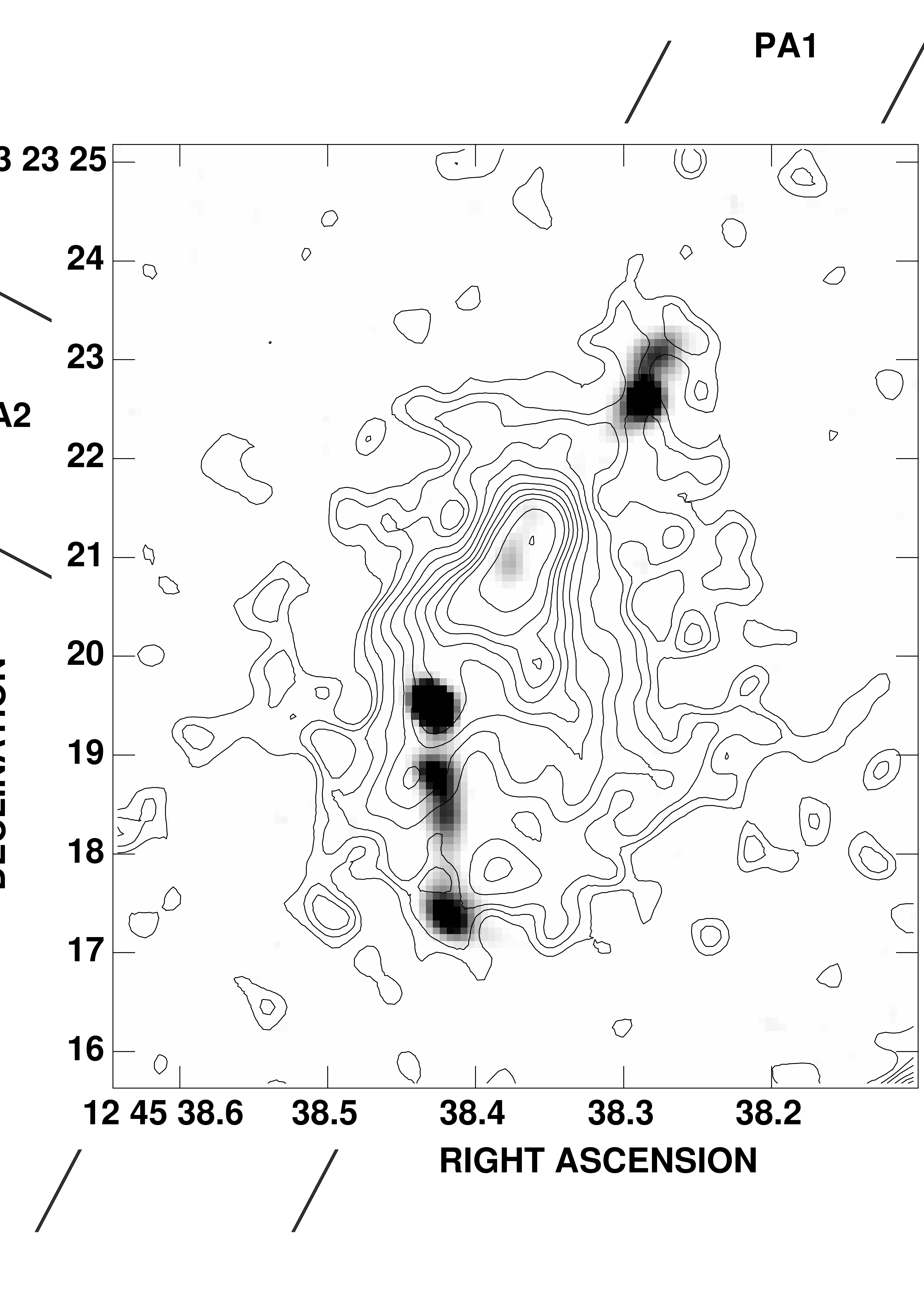,width=8cm}}
\caption{ \label{LyaVLA} The Ly$\alpha$ image of 1243+036 in contours with
the 8.3 GHz VLa map in greyscale superimposed. Radio component B1, the
bending point of the radio structure, coincides
with enhanced Ly$\alpha$ emission in one of the `arms'. Indicated around the
edges are the slit positions,
PA1 and PA2, that were used in the high resolution spectroscopy.
}
\end{figure*} \newpage \clearpage

\end{document}